\documentclass{aa}  
\usepackage{amsmath}
\usepackage{graphicx}
\usepackage{txfonts}
\usepackage{natbib}

\bibpunct{(}{)}{;}{a}{}{,} 
\usepackage{hyperref}
\hypersetup{
        colorlinks = true,
        linkcolor = blue,
        anchorcolor = red,
        citecolor = blue,
        filecolor = red,
        urlcolor = red
}

\def\ie{{\it i.e.,}\,}
\def\eg{{\it e.g.,}\,}

\def\la{\hbox{\raise.5ex\hbox{$<$} 
    \kern-1.1em\lower.5ex\hbox{$\sim$}}} 
\def\ga{\hbox{\raise.5ex\hbox{$>$} 
    \kern-1.1em\lower.5ex\hbox{$\sim$}}}

\newcommand{\dgr}{\mbox{$^\circ$}}           
\newcommand{\Msun}{\mbox{M$_\odot$\,}}         
\newcommand{\Lsun}{\mbox{$L_\odot$}}         
\newcommand{\cm}{\mbox{\ cm}}                
\newcommand{\g}{\mbox{\ g}}                  
\newcommand{\s}{\mbox{\ s}}                  
\newcommand{\K}{\mbox{\ K}}                  
\newcommand{\erg}{\mbox{\ erg }}              
\newcommand{\cms}{\mbox{\ cm s${}^{-1}$}}    
\newcommand{\Ks}{\mbox{\ K s${}^{-1}$}}    
\newcommand{\mes}{\mbox{\ m s${}^{-1}$}}    
\newcommand{\gcm}{\mbox{\ g cm${}^{-3}$}}    
\newcommand{\gcms}{\mbox{$\g\cm^{-1}\s^{-1}$}}    
\newcommand{\erggs}{\mbox{$\erg\g^{-1}\s^{-1}$}}  
\newcommand{\ergs}{\mbox{$\erg\s^{-1}$}}  
\newcommand{\ergss}{\mbox{$\erg\s^{-2}$}}  
\newcommand{\ergg}{\mbox{$\erg\g^{-1}$}}  
\newcommand{\dyncm}{\mbox{\ dyn$\cm^{-2}$}}  
\newcommand{\rads}{\mbox{\ rad$^2 \s^{-2}$\,}}  

\begin{document}
\bibliographystyle{aa}
   \title{The core helium flash revisited}
   \subtitle{III. From Pop I to Pop III stars}

   \author{M. Moc\'ak \inst{1},
           S. W. Campbell \inst{2,3},
           E. M\"uller \inst{4}
           \and K. Kifonidis\inst{4}}

   \institute{Institut d'Astronomie et d'Astrophysique, Universit\'e Libre de
              Bruxelles, CP 226, 1050 Brussels, Belgium\\
             \email{mmocak@ulb.ac.be}
         \and Departament de F\'isica i Enginyeria Nuclear, EUETIB, Universitat 
              Polit\'ecnica de Catalunya, C./Comte d'Urgell 187, E-08036 
              Barcelona, Spain\\
              \email{simon.w.campbell@upc.edu}
         \and Centre for Stellar and Planetary Astrophysics, School of 
              Mathematical Sciences, Monash University, Melbourne 3800, 
              Australia 
         \and 
              Max-Planck-Institut f\"ur Astrophysik,
              Postfach 1312, 85741 Garching, Germany\\
             }

   \date{Received  ........................... } 

 
  \abstract
   {Degenerate ignition of helium in low-mass stars at the end of the
     red giant branch phase leads to dynamic convection in their
     helium cores. One-dimensional (1D) stellar modeling of this
     intrinsically multi-dimensional dynamic event is likely to be
     inadequate. Previous hydrodynamic simulations imply that the
     single convection zone in the helium core of metal-rich Pop I
     stars grows during the flash on a dynamic timescale. This may
     lead to hydrogen injection into the core, and a double convection
     zone structure as known from one-dimensional core helium flash
     simulations of low-mass Pop III stars. }
   {We perform hydrodynamic simulations of the core helium flash in
     two and three dimensions to better constrain the nature of these
     events. To this end we study the hydrodynamics of convection
     within the helium cores of a 1.25 \Msun metal-rich Pop I star
     (Z=0.02), and a 0.85 \Msun metal-free Pop III star (Z=0) near the
     peak of the flash. These models possess single and double
     convection zones, respectively.}
   {We use 1D stellar models of the core helium flash computed with
     state-of-the-art stellar evolution codes as initial models for
     our multidimensional hydrodynamic study, and simulate the
     evolution of these models with the Riemann solver based
     hydrodynamics code Herakles which integrates the Euler equations
     coupled with source terms corresponding to gravity and nuclear
     burning.}
   {The hydrodynamic simulation of the Pop I model involving a single
     convection zone covers 27 hours of stellar evolution, while the
     first hydrodynamic simulations of a double convection zone, in
     the Pop III model, span 1.8 hours of stellar life.  We find
     differences between the predictions of mixing length theory and
     our hydrodynamic simulations.  The simulation of the single
     convection zone in the Pop I model shows a strong growth of the
     size of the convection zone due to turbulent entrainment.  Hence
     we predict that for the Pop I model a hydrogen injection phase
     (\ie hydrogen injection into the helium core) will commence after
     about 23 days, which should eventually lead to a double
     convection zone structure known from 1D stellar modeling of
     low-mass Pop III stars.  Our two and three-dimensional
     hydrodynamic simulations of the double (Pop III) convection zone
     model show that the velocity field in the convection zones is
     different from that predicted by stellar evolutionary
     calculations. The simulations suggest that the double
     convection zone decays quickly, the flow eventually being 
     dominated by internal gravity waves.}
      {}


   \maketitle
%

\section{Introduction}
\label{sect:intro}


Runaway nuclear burning of helium in the core of low mass red giant
stars leads to convective mixing and burning on dynamic time scales. One
dimensional evolutionary simulations (which assume time scales much
longer than the dynamical ones) may miss key features of this rapid
phase that could have significant effects on the further evolution of
the stars. Furthermore, 1D hydrodynamical simulations of this
intrinsically multi-dimensional event is likely to be inadequate.

Our previous hydrodynamic simulations \citep{Mocak2008, Mocak2009}
imply that a $1.25\,$\Msun solar-like star can experience injection of
hydrogen into its helium core during the canonical core helium
flash near its peak. Hydrogen injection results from the growth of the 
convection zone (which is sustained by helium burning) due to turbulent
entrainment on a dynamic timescale \citep{MeakinArnett2007}, and
probably occurs for all low-mass Pop I stars, as the properties of
their cores are similar at the peak of the core helium flash
\citep{SweigertGross1978}. An obvious consequence of this scenario is
that the convection zones are enlarged in these stars.  Whether they
fail to dredge up nuclear ash to the atmosphere shortly after the
flash is still unclear. However, such a dredge up could explain the 
Al-Mg anticorrelation found in red giants at the tip of the red giant branch
\citep{Shetrone1996a, Shetrone1996b, Yong2006}. In 1D simulations one has to
manipulate the properties of the core helium flash to achieve such a
dredge up, \eg by changing the ignition position of the helium in the
core \citep{PaczynskiTremaine1977} or by forcing inward mixing
of hydrogen \citep{Fujimoto1999}.

Canonical (1D) stellar evolution calculations predict hydrogen
injection during the core helium flash and subsequent dredge-up of
nuclear ashes to the atmosphere only for Pop III
\footnote{They are supposed to be the first stars in the Universe and
  seem to be extremely rare, as the most metal-poor star discovered up
  to now has a metallicity of [Fe/H] $\sim -5.5$ \citep{Frebel2005}.}
and extremely metal-poor (EMP; with intrinsic metallicities [Fe/H]
$\lesssim -4$) stars. This is a promising scenario for explaining the
peculiar abundances of carbon and nitrogen observed in Galactic EMP
Halo stars \citep{Campbell2008}. If these stars are assumed to be 
polluted by accretion of
CNO-rich interstellar matter they will possibly experience hydrogen
injection but no subsequent dredge-up, because a large CNO
metallicitiy (as compared to the intrinsic [Fe/H] metallicity) in the
stellar envelope influences the ignition site of the first major core
helium flash, and thereby the occurrence of the dredge-up
\citep{Hollowell1990}. The helium abundance adopted in the stellar
models also seems to influence the process of hydrogen injection
itself as shown by \citet{SchlattlCassisiSalaris2001}, while the same
authors find that the injection process seems to be independent of the
assumed convection model.

Stellar models with a higher intrinsic metallicity, \ie [Fe/H] $> -4$,
do not inject hydrogen into the helium core, and consequently there is
also no dredge-up of CNO-rich nuclear ashes to the atmosphere
\citep{FujimotoIben1990, Hollowell1990, Campbell2008}.  Whether this
is the final answer remains unclear, however, as \citet{Fujimoto1999}
with his semi-analytic study and a postulated hydrogen injection
followed by a dredge-up could show that such a scenario can explain
some peculiarities observed in the spectra of red-giant stars (related
to CNO elements and $^{24}$Mg) with metallicities as large as [Fe/H]
$\lesssim -1$.

There exist two main reasons why hydrogen injection episodes occur
only in Pop III and EMP stars: ($i$) these stars possess a flatter
entropy gradient in the hydrogen burning shell, and ($ii$) they ignite
helium far off center, relatively close to the hydrogen-rich envelope
\citep{FujimotoIben1990}. However, Pop II and Pop I stars could also
mix hydrogen into the helium core during the core helium flash
\footnote{According to \citet{SchlattlCassisiSalaris2001} the
  occurrence of a hydrogen flash is favored by a higher electron
  degeneracy in the helium core, which leads to helium ignition closer
  to the hydrogen shell. }

\begin{itemize}
\item 
  if the flash was more violent, and thus the helium convection zone
  wider \citep{DespainSalo1976, Despain1981}. This scenario is
  disfavored by the facts that the flash is less violent in stars with
  higher metallicity as less energy is needed to lift the degeneracy
  of the less massive cores \citep{SweigertGross1978}, and that helium
  ignition occurs at smaller densities \citep{FujimotoIben1990},

\item 
  or if the entropy gradient between the hydrogen and helium burning
  shell was sufficiently shallow \citep{Iben1976, Fujimoto1977}. A
  small entropy gradient would allow the convective shell in the
  helium core to reach the hydrogen layers even though the flash
  itself would not be very violent. This scenario is also disfavored
  as solutions to the stellar structure equations are robust with many
  different groups getting very similar results \ie no hydrogen
  injection \citep{FujimotoIben1990, Hollowell1990, Campbell2008} 

\item 
  or if a growth of the helium convection zone through turbulent
  entrainment at the convective boundaries \citep{Mocak2008,
    Mocak2009} could be sustained for a sufficiently long period of
  time.
\end{itemize}  

A hydrogen injection phase also occurs in low-mass metal deficient
stars on the asymptotic giant branch (AGB) at the beginning of the
thermally pulsing stage (TPAGB)
\footnote{If hydrogen injection occurs at the tip of the red giant
  branch branch (RGB), it does not occur on the AGB -- the star
  evolves like a normal thermally pulsating Pop I or II star
  \citep{SchlattlCassisiSalaris2001}}.

Hydrogen injection is found to occur in more massive stars ($M \gtrsim
1.3\,$\Msun) with low metallicity during the TPAGB \citep{Chieffi2001,
  Siess2002, Iwamoto2004}, in ``Late Hot Flasher'' stars experiencing
strong mass loss on the RGB \citep{Brown2001, CassisiSchlattl2003},
and in H-deficient post AGB (PAGB) stars. These events are referred to
with various names in the literature. Here we use the nomenclature
``dual flashes'' \citep{Campbell2008}, since they all have in common
simultaneous hydrogen and helium flashes.

\begin{figure}
\includegraphics[width=0.99\hsize]{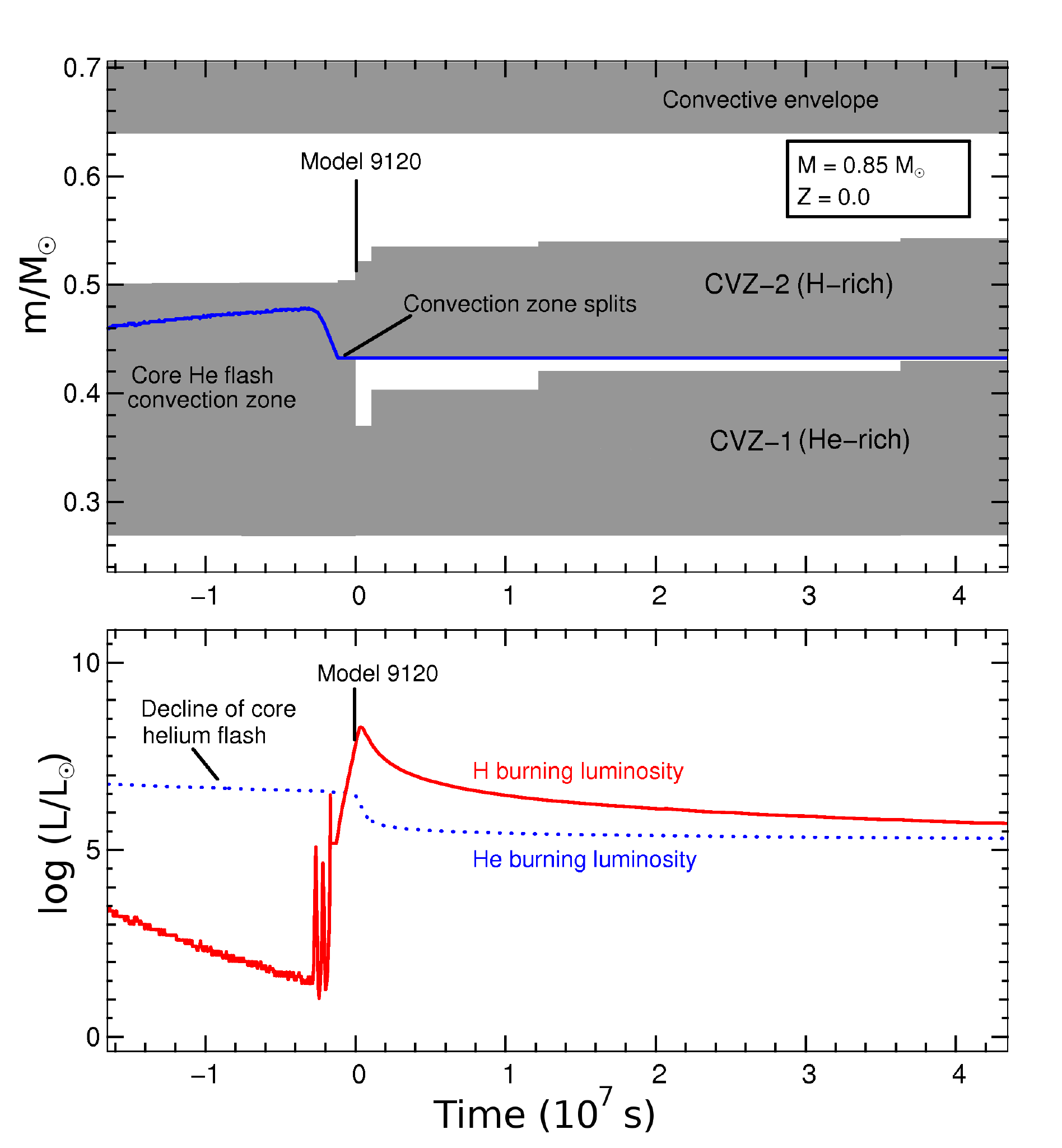}
\caption{ {\it{Upper panel:}} Kippenhahn diagram of a stellar
  evolutionary calculation during the core helium flash of a
  $0.85\,$\Msun Pop III star with convection zones sustained by helium
  (He-rich) and hydrogen (H-rich) or CNO burning (grey shaded regions,
  except for the convective envelope). The border between the helium
  and hydrogen rich layers is given by the solid blue curve. The
  location of the initial model SC (model number 9120) -
  black vertical line. {\it{Lower panel:}} the
  temporal evolution of the helium (dotted-blue) and hydrogen
  (solid-red) luminosity as a function of time. }
\label{fig.kipd}
\end{figure}

Dual flash events often lead to a splitting of the single helium
convection zone (HeCZ) into two parts (double convection zone): one
sustained by helium burning, and a second one by hydrogen burning via
CNO cycles (Fig.\,\ref{fig.kipd}). Double convection zones are
structures which are commonly encountered in stellar models, but their
hydrodynamic properties have so far only been studied for the oxygen
and carbon burning shell of a $23\,$\Msun star by \citet{Meakin2006}.

In the following we describe two-dimensional (2D) and
three-dimensional (3D) hydrodynamic simulations of a helium core
during the core helium flash with a single convection zone (Pop I; in
3D only) and a double convection zone (Pop III, in 2D and 3D),
respectively. Previous studies have indicated that there is a strong
interaction between the adjacent shells of a double convection zone by
internal gravity waves \citep{Meakin2006}. 

We introduce the stellar models used as input for our hydrodynamic
simulations in Sect.\,\ref{sect:regime}, briefly discuss the physics
included in our simulations in Sect.\,\ref{sect:input}, and give a
short description of our hydrodynamics code and the computational
setup in Sect.\,\ref{sect:hcode}. Subsequently, we present and compare
the results of our 2D and 3D hydrodynamic simulations in
Sect.\,\ref{sect:tmpevol}.  In particular, we discuss turbulent
entrainment at the convective boundaries for our single convection
zone model, the temporal evolution of its kinetic energy density, and
how our results compare with the predictions of mixing-length theory
(MLT). We proceed similarly for our hydrodynamic double convection
zone model, except for turbulent entrainment for reasons which become
clear later.  Finally, a summary of our findings is given in
Sect.\,\ref{sect:sum}.

\section{Physical Conditions and Initial Data}
\label{sect:regime}

\begin{figure*}
\includegraphics[width=0.49\hsize]{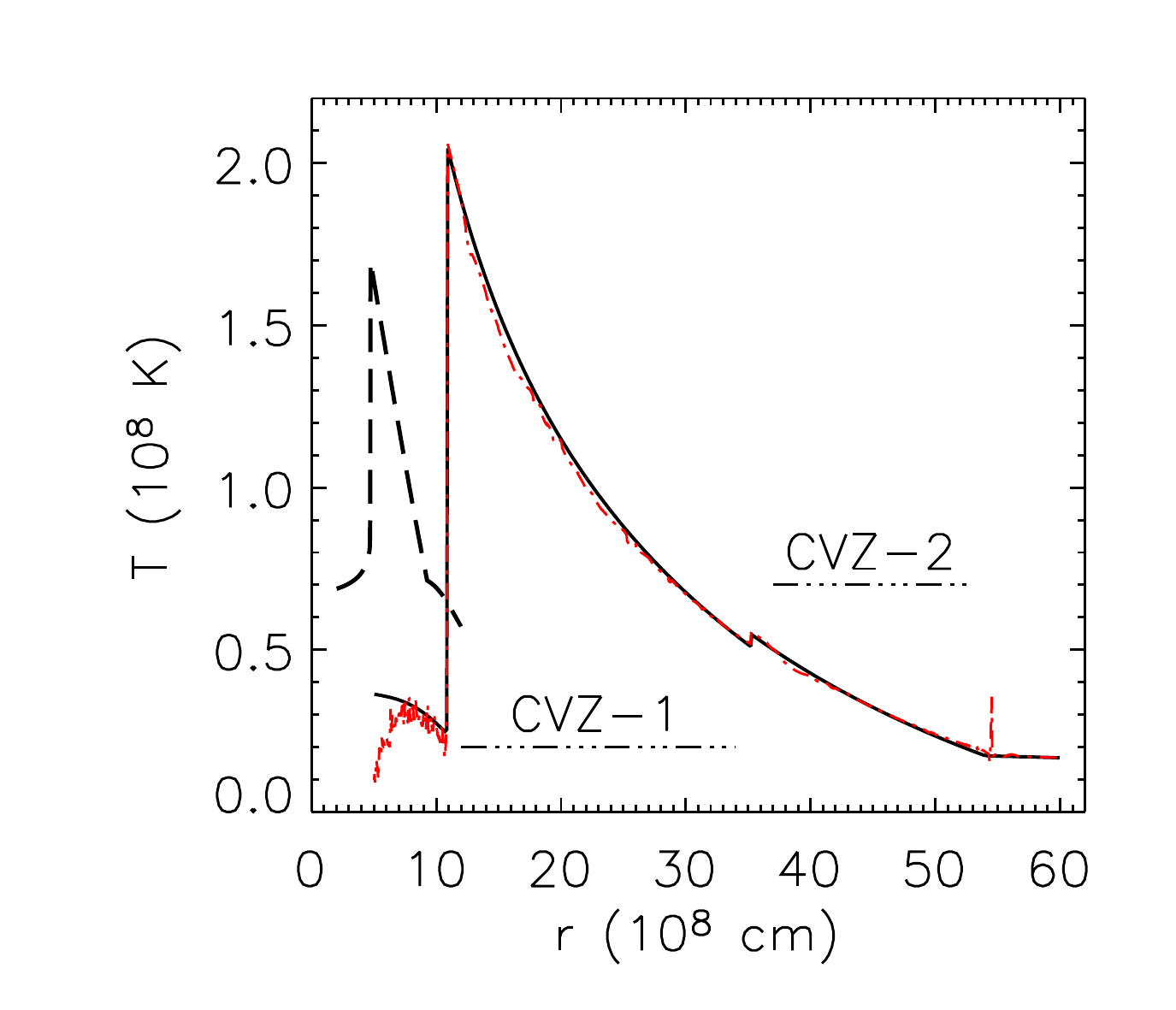}
\includegraphics[width=0.49\hsize]{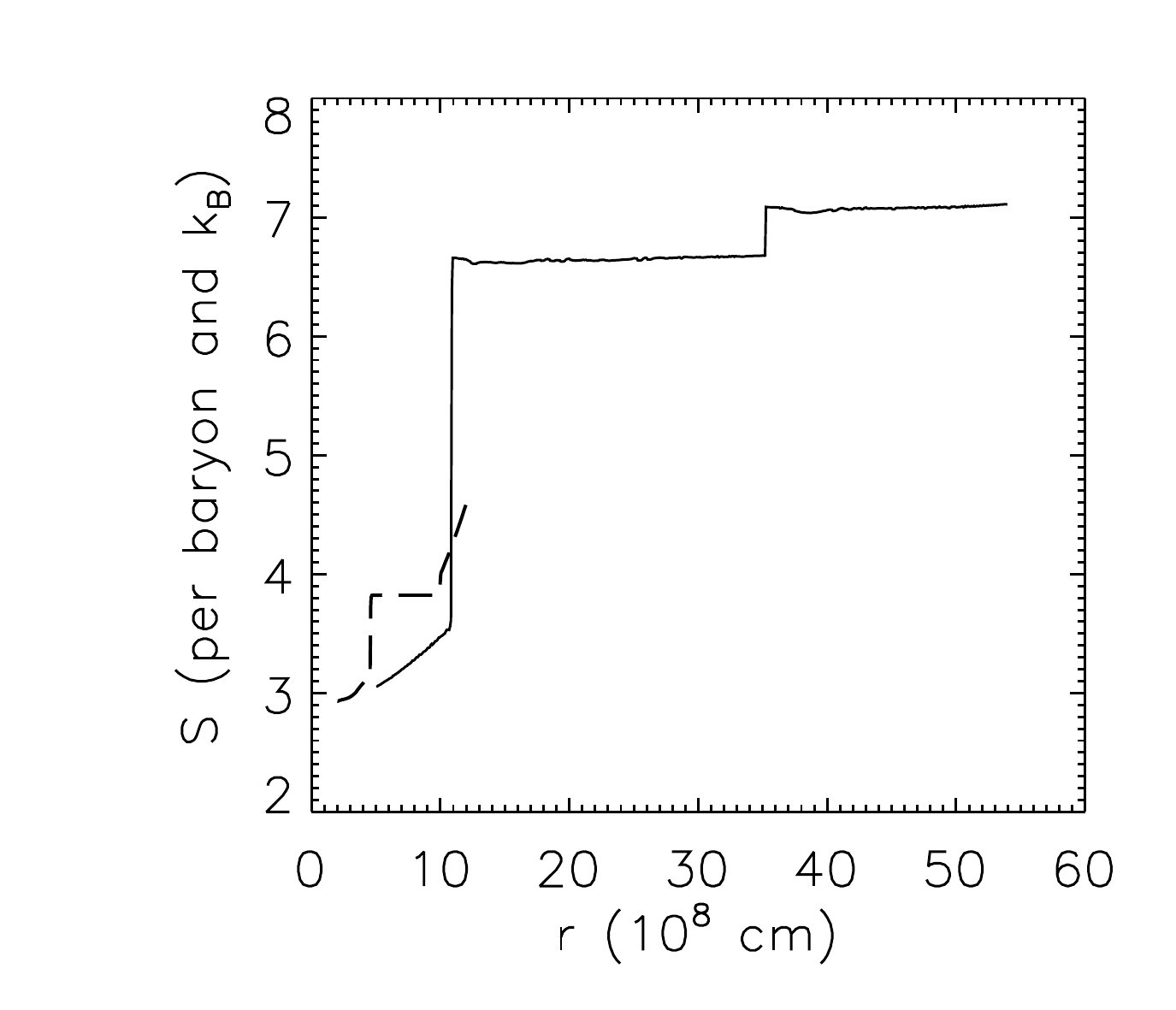}
\caption{ {\it{Left:}} Temperature distribution in the helium core in
  model M (long-dashed), and in model SC (solid) with its stabilized
  counterpart (dash-dotted red), respectively. The two parts of the
  double convection zone present in model SC are denoted by CVZ-1 and
  CVZ-2, respectively.  {\it{Right:}} Entropy distribution of model M
  (solid) and model SC (long-dashed), respectively.  }
\label{fig.inimt}
\end{figure*}

Our initial helium core models (Tab.\,\ref{tab.inim}) with single and
double convection zones (models M and SC, respectively) are obtained
from 1D stellar evolutionary calculations of a Pop I star (Z=0.02) with a
mass of $1.25\,$\Msun, and a Pop III star (Z=0) with a mass of 0.85
\Msun, 
\footnote{Metal-free stars with masses $ \gtrsim 1\,$\Msun ignite
  helium at the center before electrons become degenerate
  \citep{Fujimoto2000}.}, 
respectively. Both models are characterized by an off-center helium
ignition which results in convection zones characterized by a
temperature gradient close to the adiabatic
\footnote{The entropy $S$ (Fig.\,\ref{fig.inimt}) and the degeneracy
  parameter $\psi$ remain almost constant in the convection zones,
  which is a result of the almost adiabatic temperature gradient, \ie
  the temperature relation $T \propto \rho^{\gamma - 1}$ with the
  adiabatic exponent $\gamma \sim 5/3$ holds. Since $\rho T^{-3/2} =
  f(\psi)$, the degeneracy parameter $\psi$ is constant.}  
one above the helium burning source.

The helium cores of both models are composed of a gas which is almost
completely ionized, as the ionization potentials of both He and He$^+$
($E_B \sim 24.7\,$eV and $E_B \sim 54.4\,$eV, respectively) are very
small compared to the thermal energy, \ie
\begin{eqnarray}
  \frac{E_B}{k_B T} < 10^{-2} \, ,
\end{eqnarray}  
where $k_B$ is Boltzmann's constant, and $T > 4.7 \times 10^7\,$K ($T
> 1.7 \times 10^7\,$K) holds for the temperature inside the helium
core of model M (SC) which has a radius $r \sim 1.9 \times 10^9\,$cm
($r \sim 5.4 \times 10^9\,$cm ).

In the central part of the models (beneath the convection zones) the
electron density is so high that the gas is highly degenerate.  On the
other hand, the density of electrons in the single and double
convection zone is much lower due to a strong expansion that occurred a
little earlier in the evolution. Thus, the degeneracy has been lifted
in the convection zones, \ie the ratio of the Fermi energy $E_{F,E}$
of the electrons \citep{CoxGiuli2004} and their typical thermal energy
is small,
\begin{eqnarray}
  \frac{E_{F,E}}{k_B T} \sim 2.47 \,\,\,\,\, (0.14)   
\end{eqnarray}
where $E_{F,E} \sim 4.1 \times 10^{-8} \erg$ ($E_{F,E} \sim 1.2 \times
10^{-9}\erg$ ) in the middle of the convection zone of model M (SC) at
a radius $r \sim 7 \times 10^8$\,cm ($r \sim 3.3 \times 10^9\,$cm).

The ions can be described as an ideal, non-relativistic Boltzmann gas,
because the ratio of their Fermi energy $E_{F,I}$ and their typical
thermal energy is small,
\begin{eqnarray}
  \frac{E_{F,I}}{k_B T} \sim 2\times 10^{-4} \,\,\,\,\, (1\times 10^{-5}) 
\end{eqnarray}
where $E_{F,I} \sim 3.4 \times 10^{-12} \erg$ ($E_{F,I} \sim 1 \times
10^{-13}\erg$ ) in the middle of the convection zone of model M (SC)
at a radius $r \sim 7 \times 10^8 \cm$ ($r \sim 3.3 \times 10^9\,$cm).
Coulomb forces between the ions are negligible, too, as the Coulomb
energy corresponding to the mean ion-ion distance is less than 30\%
(10\%) of the thermal energy of the ions for model M (SC) in the
middle of the convection zone.

\begin{table*} 
\caption[Initial models:~Properties of M \& SC] {Some properties of
  the initial models M \& SC: Total mass $M$, stellar population,
  metal content $Z$, mass $M_{He}$ and radius $R_{He}$ of the helium
  core ($X(^{4}He) > 0.98$), ratio of the binding energy of the
  electrons $E_B$ in the helium core and of their thermal energy
  $k_{B} T$, ratio of the Coulomb and thermal energy of the ions
  $\Gamma_{cnv}$, electron degeneracy parameter in the convection zone
  $\Psi_{cnv}$, nuclear energy production rate due to helium $L_{He}$
  and hydrogen $L_{He}$ burning, temperature maximum $T_{max}$, and
  radius $r_{max}$ and density $\rho_{max}$ at the temperature
  maximum.}
\vspace{0.4cm}

\begin{tabular}{l|lllll|lll|lllll} 
\hline
\hline
Model & $M$       & Pop. & $Z$ & $M_{He}$  & $R_{He}$ & E$_B$/k$_B$T & 
$\Gamma_{cnv}$ & $\Psi_{cnv}$ & $L_{He}$   & $L_{H}$   & $T_{max}$  & $r_{max}$   
& $\rho_{max}$ \\ 
       & $[\Msun]$ &      &     & $[\Msun]$ & $[10^9\cm]$ & & &  &
 $[10^9\Lsun]$  & $[10^9\Lsun]$ & $[10^8\K]$ & $[10^8\cm]$ & $[10^5\gcm]$  \\
\hline
M  & $1.25$ & I & $0.02$ & $0.47$ & $1.91$ & $<$ 0.006 & 0.3 & ~5 & $1.03$ 
  & $10^{-6}$ & $1.70$ & $4.71$ & $3.44$ \\
\hline 
SC  & 0.85 & III & $0.00 $ & $0.5$ & $5.45$  & $<$ 0.016 & 0.1 & -2 & $0.004$  
  & 0.07 & $2.04$ & $11.$ & $0.08$  \\
\hline
\end{tabular} 
\label{tab.inim} 
\end{table*} 

Convection may become turbulent showing random spatial and temporal
fluctuations. This can be characterized by the dimensionless Reynolds
number $R_e$ \citep{LandauLifshitz1966} which is basically the ratio
of inertial to viscous forces. The turbulent regime is entered once
$R_e$ exceeds a certain critical value $R_{crit}$, typically being of
the order of $10^3$, at which small fluctuations in the flow are no
longer damped. We estimate that the Reynolds numbers in the central
convection zones of our models are
\begin{eqnarray}
  R_e \sim \frac{v \cdot l \cdot \rho}{\eta} \sim 10^{14} \,\,\,\,\, (10^{14})
\end{eqnarray}
where $\rho \sim 10^5 \gcm$ ($\rho \sim 10^3 \gcm$), $l \sim 10^8 \cm$
($l \sim 10^9 \cm$), $v \sim 10^6 \cms$ ($v \sim 10^5 \cms$), and
$\eta \sim 10^{5} \gcms$ ($\eta \sim 10^{3} \gcms$) are the typical
densities, lengths, velocities, and viscosities 
\footnote{The estimate of $\eta$ for a strongly degenerate and
  completely ionized helium gas is based on the formula of
  \citet{Itoh1987}.}
of the convective flow in model M (SC) as predicted by stellar
evolutionary calculations.  These values imply that the flow is highly
turbulent, which leads to complications when trying to simulate such
flows, as turbulence is an intrinsically three-dimensional phenomenon
involving a large range of spatial and temporal scales. We recall
that, in three-dimensional turbulent flow, large structures are
unstable and cascade into smaller vortices down to molecular scales
where the kinetic energy of the flow is eventually dissipated into
heat.

If $L$ is the largest (integral) scale characterizing a flow, and $l$
the scale where viscous dissipation dominates, one has the well known
relation:
\begin{eqnarray}
  \frac{L}{l} \sim R^{3/4}_{e} 
\label{eq2.11}
\end{eqnarray}
In the convection zone, where the Reynolds number $R_{e} \sim
10^{14}$, one finds $L/l \sim 10^{10.5}$.  Therefore, the number of
grid points $N$ that a numerical simulation would require to resolve
all relevant length scales is $N \sim (L/l)^3 \sim R^{9/4}_{e} \sim
10^{31.5}$, which is roughly 22 orders of magnitude larger than the
highest resolutions that can be handled by present day computers. 

To account for turbulence on the numerically unresolved scales, one
usually adopts sub-grid scale models \eg the quite popular one by
\citet{Smagorinsky1963}, which describe the energy transfer from the
smallest numerically resolved turbulent elements to those at the
dissipation length scale using various (phenomenological and/or
physical) model and flow dependent parameters. Here we have not 
adopted a sub-grid scale model, however it has been shown that if one 
does or does not use such a model it seems not to lead to qualitative 
differences in the hydrodynamic behavior of the core helium flash 
\citep{Achatz1995}.

\subsection{Initial stellar model M}
\label{subs:inim}

The initial model M (Tab.\,\ref{tab.inim}, Fig.\,\ref{fig.inimt}) was
obtained with the stellar evolution code GARSTEC
\citep{WeissSchlattl2000, WeissSchlattl2007} by Achim Weiss, and
represents a 1.25\,\Msun star at the peak of the core helium flash
characterized by an off-center temperature maximum at the base of a
single convection zone sustained by helium burning.  Additional
information about the model can be found in \citet{Mocak2008,
  Mocak2009}.

As we are interested here only in the hydrodynamic evolution of the
helium core
\footnote{The helium core is basically a white dwarf sitting inside a
  red giant star. It has a relatively small radius -- comparable to
  that of the Earth -- but contains almost half of the total mass of
  the star (Tab.\,\ref{tab.inim}).}, 
we consider of model M only the shell between $r = 2 \times 10^8\,$cm
to $r = 1.2\times10^9\,$cm, which contains the single convection zone
powered by triple-$\alpha$ burning. Originally, the convection zone
reaches from $4.72 \times 10^8\,$cm (local pressure scale height $2.9
\times 10^8\,$cm) to $9.2 \times 10^8\,$cm (local pressure scale
height $1.4 \times 10^8\,$cm).  From the bottom to the top of the
convection zone the pressure changes by $\sim 1$ order of magnitude,
from $6.6 \times 10^{21} \dyncm$ to $7.1\times 10^{20} \dyncm$.  We
note that both the base and the top of the convection zone are located
sufficiently far away from the (radial) grid boundaries.

\begin{figure*}
\includegraphics[width=0.49\hsize]{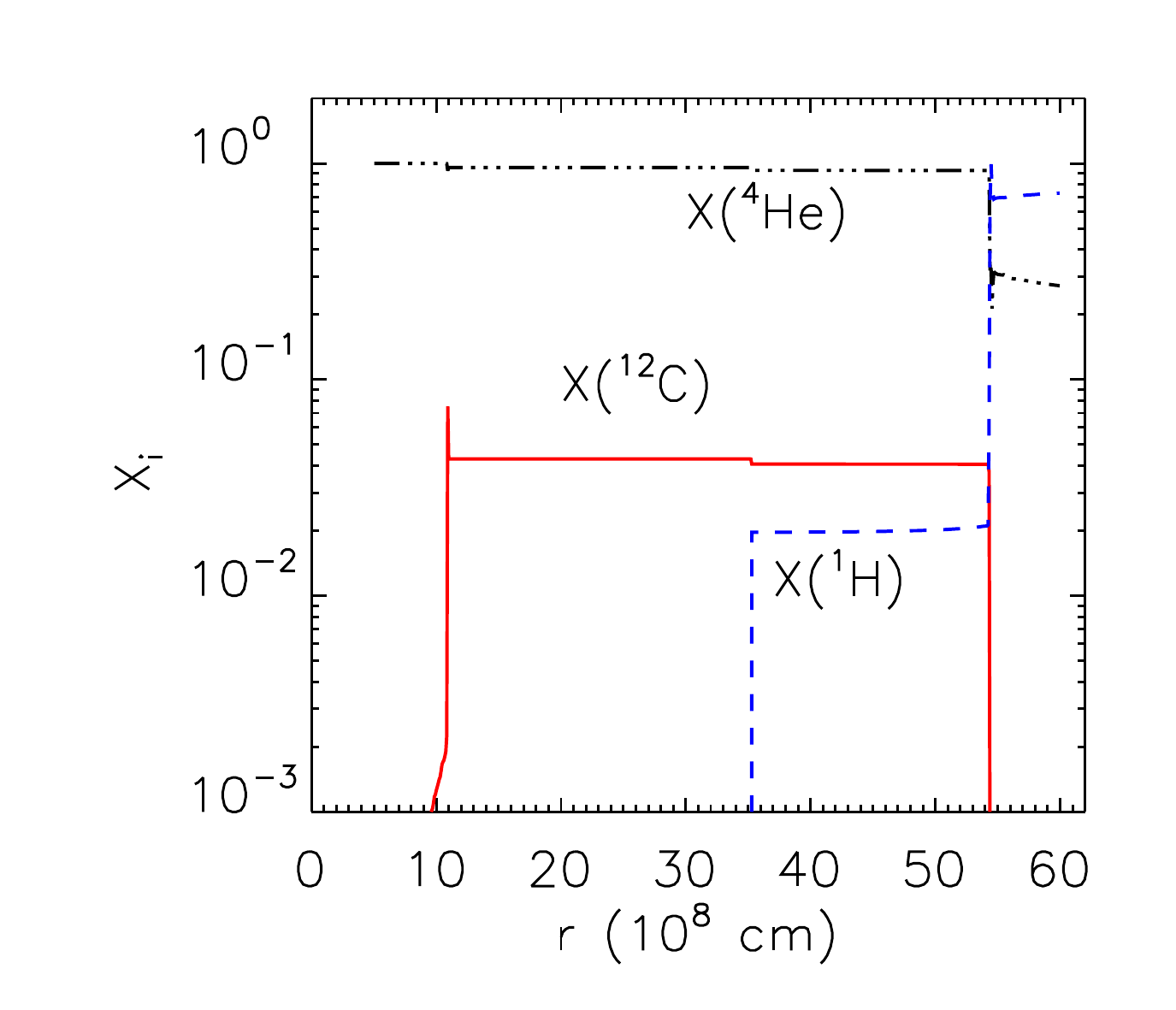} 
\includegraphics[width=0.49\hsize]{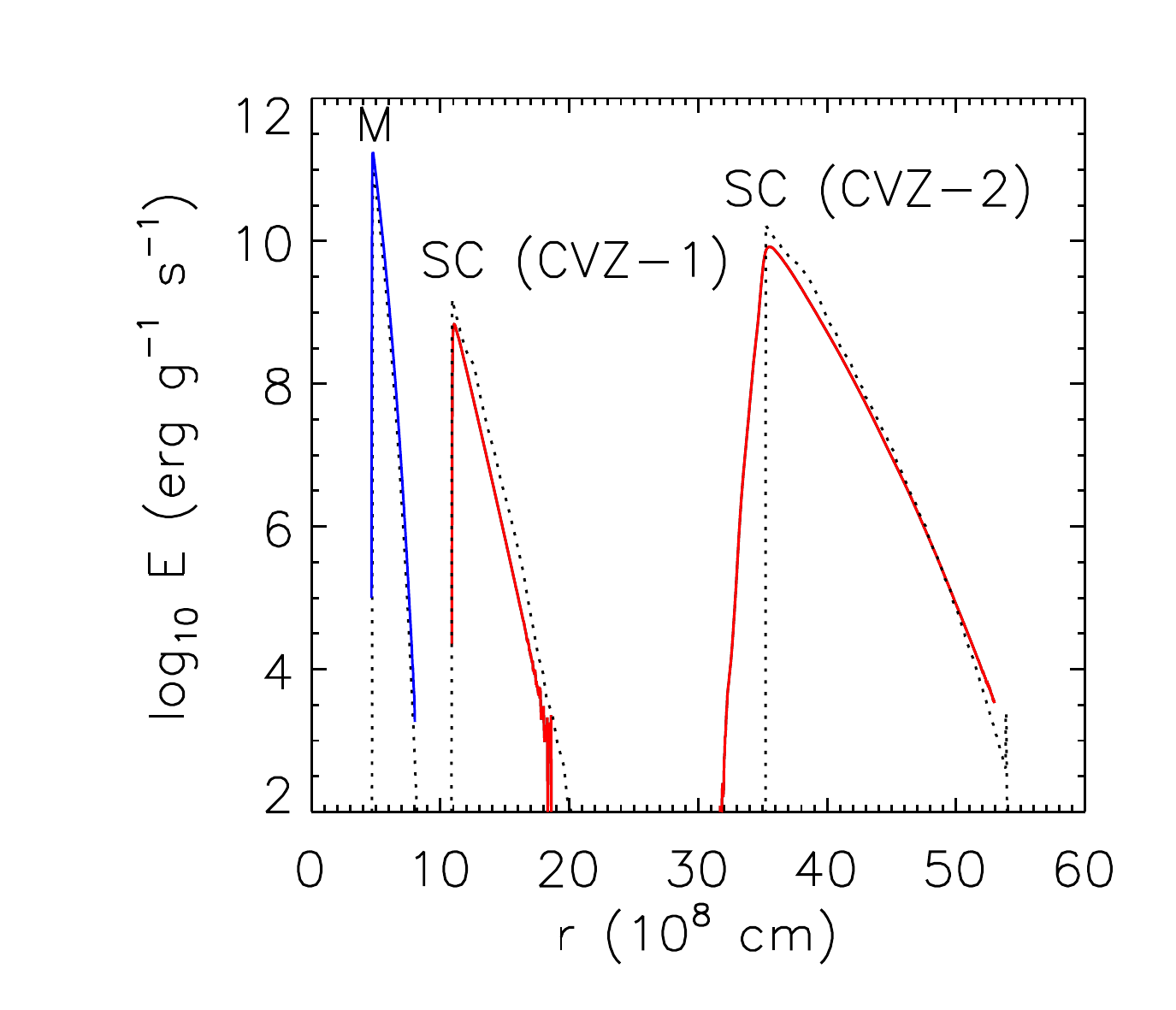}
\caption{ {\it{Left:}} Chemical composition of the helium core in
  model heflpopIII.2d.2 (SC).  {\it{Right:}} Nuclear energy production
  rate as a function of radius $r$. Initial rates (at t=0) are indicated 
  by dotted-black curves. Rates in model heflpopI.3d (SC) at $t =
  6400\,$s (solid-red), and in model M at $t \sim 10^5\,$~s
  (solid-blue), respectively.}
\label{fig.inicomp}
\end{figure*}

Model M contains the chemical species $^{1}$H, $^{3}$He, $^{4}$He,
$^{12}$C, $^{13}$C, $^{14}$N, $^{15}$N, $^{16}$O ,$^{17}$O,$^{24}$Mg,
and $^{28}$Si.  However, since we are not interested in details of its
chemical evolution, we considered only the abundances of $^{4}$He,
$^{12}$C, and $^{16}$O in our hydrodynamic simulations. This is
justified as the triple $\alpha$ reaction dominates the nuclear energy
production rate during the core helium flash. For the remaining
composition we assume that it can be represented by a gas with a mean
molecular weight equal to that of $^{20}$Ne, as its nucleon number agrees
well with the average nucleon number of the neglected nuclear species.

The stellar model had to be relaxed into hydrostatic equilibrium after
it was mapped to the numerical grid of our hydrodynamics code. This
was achieved with an iterative procedure, which keeps the density
distribution of the model almost constant, but modifies its pressure
distribution to achieve hydrostatic equilibrium
\citep{Mocak2009phd}. This mapping process has a negligible effect on
the stellar structure.

\subsection{Initial stellar model SC}
\label{subs:inisc}

The initial model SC (Tab.\,\ref{tab.inim}, Fig.\,\ref{fig.kipd} to
\ref{fig.inicomp}) was computed by Simon W.\,Campbell using the
Monash/Mount Stromlo stellar evolution code (MONSTAR)
\citep{Campbell2008, Wood1981}. It corresponds to a metal-free Pop III
star with a mass of 0.85\,$\Msun$ near the peak of the core helium
flash. Metal-free stars with masses $\gtrsim 1~ \Msun$ do not undergo
the core helium flash (as opposed to $M \lesssim 2.2\,\Msun$ at solar
metallicity). The helium core flash commences with a very off-center
ignition of helium in a relatively dense environment under degenerate
conditions, and results in a fast growing convection zone powered by
helium burning that relatively quickly reaches the surrounding
hydrogen shell \citep{FujimotoIben1990}. This causes sudden mixing of
protons down into the hot helium core (Fig.\,\ref{fig.inicomp}), and
leads to rapid nuclear burning via the CNO cycle \ie a hydrogen flash.
Since the core helium flash is still ongoing we refer to this as a
``Dual Core Flash'' (DCF).  We note that this event has also been
referred to as ``helium flash induced mixing''
\citep{SchlattlCassisiSalaris2001, CassisiSchlattl2003,
  WeissSchlattl2004}, and ``helium flash-driven deep mixing''
\citep{Suda2004}.  The CNO burning leads to an increase of the
temperature inside the helium-burning driven convection zone, and
causes it to split into two.  The result is a lower convection zone
still powered by helium burning and a second one powered by the CNO
cycle (Fig.\,\ref{fig.inicomp}, [right panel]). In the following we will 
refer to the split convection zone as a double convection zone.

The electron degeneracy in the double convection zone has already been
significantly lifted to $\psi \sim -2$. It can be
shown that for a degeneracy parameter $\psi < -2$, the gas pressure is
essentially that of a nondegenerate gas \citep{Clayton1968}. This
confirms our previous conclusion based on the ratio of the Fermi and
thermal energy of the electrons (Sect.\,\ref{sect:regime}).

In our hydrodynamic simulations we considered a shell from model SC
which extends from $r= 5 \times 10^8\,$cm to $r=6 \times10^9\,$cm
containing the double convection zone. Initially, the inner convection
zone (powered by triple-$\alpha$ burning) covers a region from $r
\approx 11 \times 10^8\,$cm (local pressure scale height $6 \times
10^8\,$cm) to $r \approx 35 \times 10^8\,$cm (local pressure scale
height $7 \times 10^8\,$cm), while the outer convection zone stretches
from there up to $54 \times 10^8\,$cm (local pressure scale height $5.2 \times
10^8\,$cm). From the bottom to the top of the double convection zone
the pressure changes by $\sim 3$ orders of magnitude from $1 \times
10^{20} \dyncm$ erg to $1.4 \times 10^{17} \dyncm$.  Again, we have
ensured that the region of interest was located sufficiently far away
from the radial grid boundaries.

Our hydrodynamic simulations were performed adopting the mass fractions
of all the species used in the corresponding stellar evolutionary
calculations, namely $^1$H, $^3$He, $^4$He, $^{12}$C, $^{14}$N, and
$^{16}$O. Since the evolutionary model did not include $^{13}$C and
$^{13}$N, we determined their mass fractions assuming that the CNO
cycle had been operating in equilibrium. The remaining composition is
represented by a gas with the molecular weight of $^{20}$Ne.

The model was relaxed to hydrostatic equilibrium in the same manner as
in case of model M. This process resulted in small fluctuations in the
temperature profile (Fig.\,\ref{fig.inimt}), which were smeared out
after the onset of convection.

\section{Input physics}
\label{sect:input}

The input physics of our hydrodynamic simulations is identical to that
one described in \citet{Mocak2008}, except for the number of nuclear
species employed in the simulations based on the initial model SC. We
use an equation of state that includes contributions from radiation,
ideal Boltzmann gases, and an electron-positron component
\citep{TimmesSwesty2000}.  Thermal transport was neglected as the
maximum amount of energy transported by radiation and heat conduction
is smaller than the convective flux by at least seven (three) orders
of magnitude in model M (SC).  Neutrinos act as a nuclear energy sink,
but carry away less than $< 10^2 \erggs$. This is a negligible amount
(especially for the timescales covered by our simulations) when
compared to the maximum nuclear energy production $\dot{\epsilon}$,
which is $\sim 10^{11} \erggs$ for model M, and $\sim 10^{10} \erggs$
in model SC, respectively (Fig.\ref{fig.inicomp}).

\subsection{Nuclear reactions}

We employed two different nuclear networks for our simulations, as the
nuclear species considered in models M and SC differ.  The nuclear
reaction network used in the hydrodynamic simulation based on the
initial model M (Tab.\,\ref{tab.inim}) consists of four nuclei
(Sect.\,\ref{subs:inim}) coupled by seven reactions. The network is
identical to that one described by \citet{Mocak2008} \ie
\begin{flushleft}
\begin{tabular}{lclclclcll} 
He$^{ 4}$ &+& C$^{12}$  &$\rightarrow$& O$^{16}$  &+& $\gamma$ & &\\
He$^{ 4}$ &+& O$^{16}$  &$\rightarrow$& Ne$^{20}$ &+& $\gamma$ & &\\
O$^{16}$  &+& $\gamma$ &$\rightarrow$& He$^{ 4}$ &+& C$^{12}$  & &\\ 
Ne$^{20}$ &+& $\gamma$ &$\rightarrow$& He$^{ 4}$ &+& O$^{16}$  & &\\
C$^{12}$  &+& C$^{12}$  &$\rightarrow$& Ne$^{20}$ &+& He$^{ 4}$ & &\\
He$^{ 4}$ &+& He$^{4}$  &+& He$^{4}$  &$\rightarrow$& C$^{12}$ &+& $\gamma$ \\
C$^{12}$  &+& $\gamma$ &$\rightarrow$& He$^{4}$  &+& He$^{ 4}$ &+& He$^{4}$ 
\end{tabular}
\end{flushleft}

The nuclear reactions considered in the hydrodynamic simulations based
on the initial model SC (Tab.\,\ref{tab.inim}) are described by a
reaction network consisting of nine nuclei (Sect.\,\ref{subs:inisc})
coupled by the following 16 reactions:
\begin{flushleft}
\begin{tabular}{lclcllllllll} 
H$^{1}$   &+& He$^3$   &$\rightarrow$& He$^{4}$  &+& $\gamma$ & & & &\\ 
He$^{4}$  &+& C$^{12}$  &$\rightarrow$& O$^{16}$  &+& $\gamma$ & & & &\\ 
He$^{4}$  &+& N$^{13}$  &$\rightarrow$& H$^{1}$   &+& O$^{16}$ & & & &\\ 
H$^{1}$   &+& C$^{13}$  &$\rightarrow$& N$^{14}$  &+& $\gamma$ & & & &\\ 
H$^{1}$   &+& C$^{12}$  &$\rightarrow$& N$^{13}$  &+& $\gamma$ & & & &\\ 
H$^{1}$   &+& O$^{16}$  &$\rightarrow$& He$^{4}$  &+& N$^{13}$ & & & &\\ 
He$^{4}$  &+& O$^{16}$  &$\rightarrow$& Ne$^{20}$ &+& $\gamma$ & & & &\\ 
C$^{12}$  &+& C$^{12}$  &$\rightarrow$& He$^{4}$  &+& Ne$^{20}$ &+ &$\gamma$ & &\\ 
N$^{13}$  &+& $\gamma$ &$\rightarrow$& H$^{ 1}$  &+& C$^{12}$ & & & &\\ 
N$^{14}$  &+& $\gamma$ &$\rightarrow$& H$^{ 1}$  &+& C$^{13}$ & & & &\\ 
O$^{16}$  &+& $\gamma$ &$\rightarrow$& He$^{4}$  &+& C$^{12}$ & & & & \\ 
Ne$^{20}$ &+& $\gamma$ &$\rightarrow$& He$^{4}$  &+& O$^{16}$ & & & &\\ 
C$^{12}$  &+& $\gamma$ &$\rightarrow$& He$^{4}$  &+& He$^{4}$ &+& He$^{4}$ & &\\
He$^{4}$  &+& He$^{4}$ &+& He$^{4}$  &$\rightarrow$& C$^{12}$ &+& 
                                                               $\gamma$ & &\\
He$^{3}$ &+& He$^{3}$ &$\rightarrow$& H$^{1}$ &+& H$^{1}$ &+ &He$^4$ &+
& $\gamma$\\ 
H$^{1}$ &+& H$^{1}$ &+& He$^{4}$  &$\rightarrow$& He$^{3}$ &+& He$^{3}$ &+&
                                                               $\gamma$ & \\
\end{tabular}
\end{flushleft}
This network reproduces the nuclear energy generation rate of the
original stellar model very well (Fig.\,\ref{fig.inicomp}).

Note, that although the value of the temperature maximum, $T_{max}$,
is higher in model SC than in model M, the energy generation rate is
lower at $T_{max}$ in model SC, because the helium mass fraction
$X(^{4}$He) is smaller in that model (0.956 as compared to 0.970 for
model M).

\section{Hydrodynamic code and computational setup}
\label{sect:hcode}
%
We use the hydrodynamics code Herakles \citep{Kifonidis2003,
  Kifonidis2006, Mocak2008, Mocak2009} which solves the Euler
equations coupled with source terms corresponding to gravity and
nuclear burning.  The hydrodynamic equations are integrated with the
piecewise parabolic method of \citet{ColellaWoodward1984} and a
Riemann solver for real gases according to
\citet{ColellaGlaz1984}. The evolution of the nuclear species is
described by a set of additional continuity equations
\citep{PlewaMueller1999}. Self-gravity is implemented according to
\citet{MuellerSteimnetz1995} and nuclear burning is treated by the
semi-implicit Bader-Deuflhard scheme \citep{Press1992}.

We performed one 3D simulation based on the initial model M, which
covered roughly $27\,hrs$ of stellar evolution
(Tab.\,\ref{modpop1tab}).  This model (henceforth heflpopI.3d) was
evolved on a computational grid consisting of a $30\dgr$-wide wedge in
both $\theta$ and $\phi$-direction centered at the equator. The small
angular size of the grid allowed us to achieve a relatively high
angular resolution ($1\dgr$) with a modest number of angular zones
($N_\phi N_\theta =30$).

\begin{figure}
\includegraphics[width=0.99\hsize]{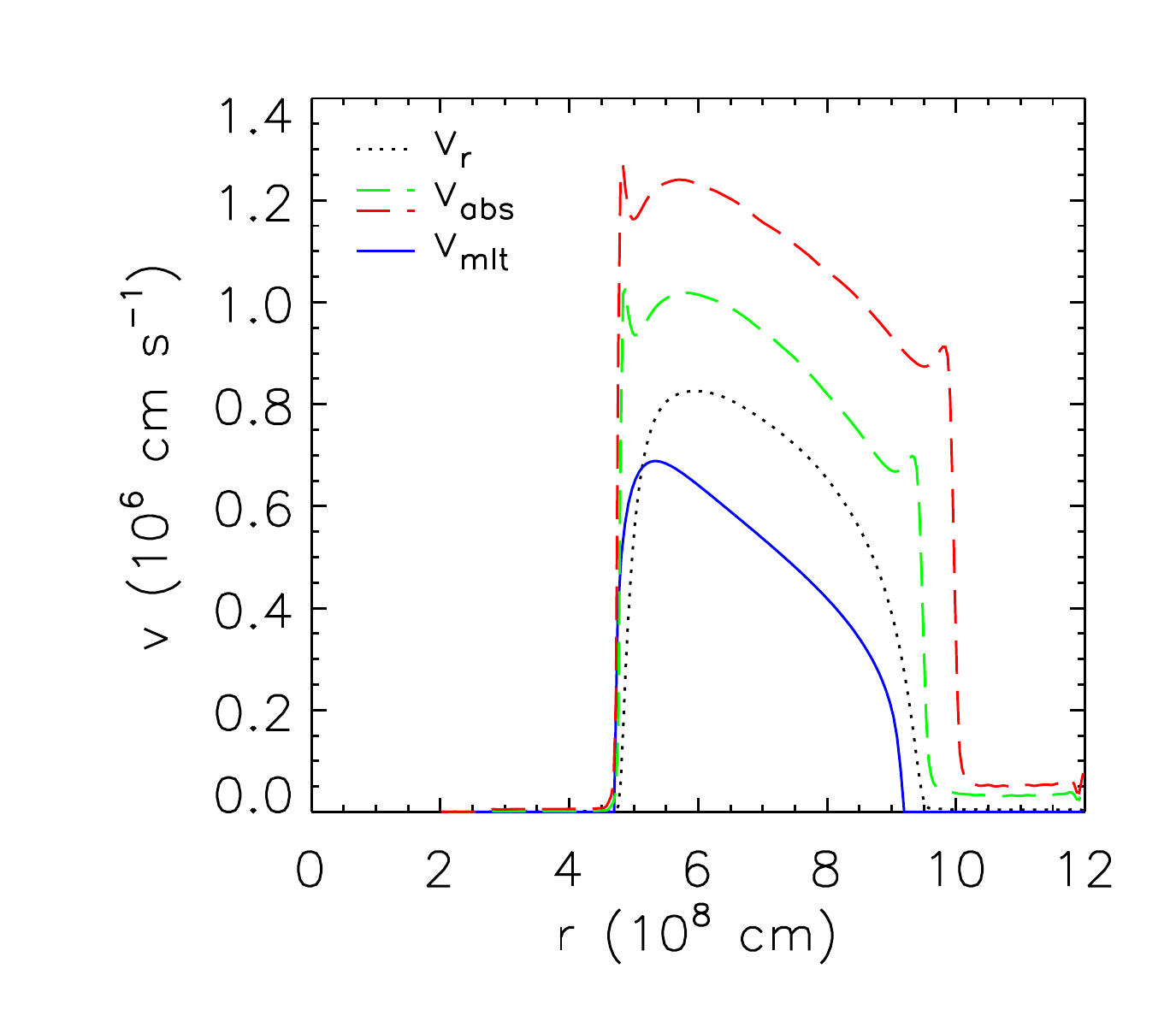}
\caption{Radial velocity distributions for the 3D model heflpopI.3d.
  The dotted and green dashed lines show the time (from 10\,000\,s to
  30\,000\,s) and angle-averaged radial velocity, $v_r$, and velocity
  modulus, $v_{abs}$, respectively. The red dashed line shows again
  the latter velocity, but time-averaged from 80\,000\,s to
  99\,000\,s. The velocity predicted by the mixing-length theory
  ($v_{mlt}$) for the initial model M is shown by the solid blue
  line. }
\label{fig.veloc}
\end{figure}

\begin{table*} 
\caption{Some properties of the 3D simulation based on model M: number
  of grid zones, wedge size $w$, grid resolution in $r$, $\theta$ and
  $\phi$ direction, estimated Reynolds number $R_e$
  \citep{PorterWoodward1994}, characteristic velocity $v_{c}$ of the
  convective flow, typical convective turnover timescale $\tau_{cnv}$,
  and maximum evolutionary time $t_{max}$.}
\begin{center}
\begin{tabular}{p{2.cm}|p{2.cm}p{1.cm}p{1.3cm}p{1.0cm}p{1.0cm}p{1.0cm}
p{1.5cm}p{1.5cm}p{1.5cm}} 
\hline
\hline
run & grid & $w$ & $\Delta r$ & $\Delta\theta$ & $\Delta\phi$ & R$_e$ &  
$v_{c}$  & $\tau_{cnv}$ & $t_{max}$ \\
$\mbox{[name]}$ & N$_r \times$N$_\theta \times$N$_\phi$ & [$\dgr$] & 
[10$^{6}$ cm] & [$\dgr$] & [$\dgr$]  & & [10$^{6}\,\cms$] & [s] & [s] \\
\hline 
heflpopI.3d & $270\times30\times30$ & 30 & 3.7 & 1 & 1 & $10^2$
 & 1.1 & 1000 & 100000 \\
\hline
\end{tabular} 
\end{center}
\label{modpop1tab} 
\end{table*}  

In addition, we performed two 2D (henceforth models heflpopIII.2d.1
and heflpopIII.2d.2) and one 3D simulation (henceforth model
heflpopIII.3d) based on the initial model SC covering about $1.8\,hrs$
and $0.39\,hrs$ of stellar evolution, respectively
(Tab.\,\ref{modpop3tab}). We used a computational grid consisting of a
$50\dgr$-wide angular wedge centered at the equator in case of models
heflpopIII.2d.1 and heflpopIII.3d, and of a $120\dgr$ wedge in case of
model heflpopIII.2d.2.

We imposed reflective boundary conditions in the radial direction and
transmissive ones in the angular directions in all our
multi-dimensional simulations.

\begin{figure}
\includegraphics[width=0.99\hsize]{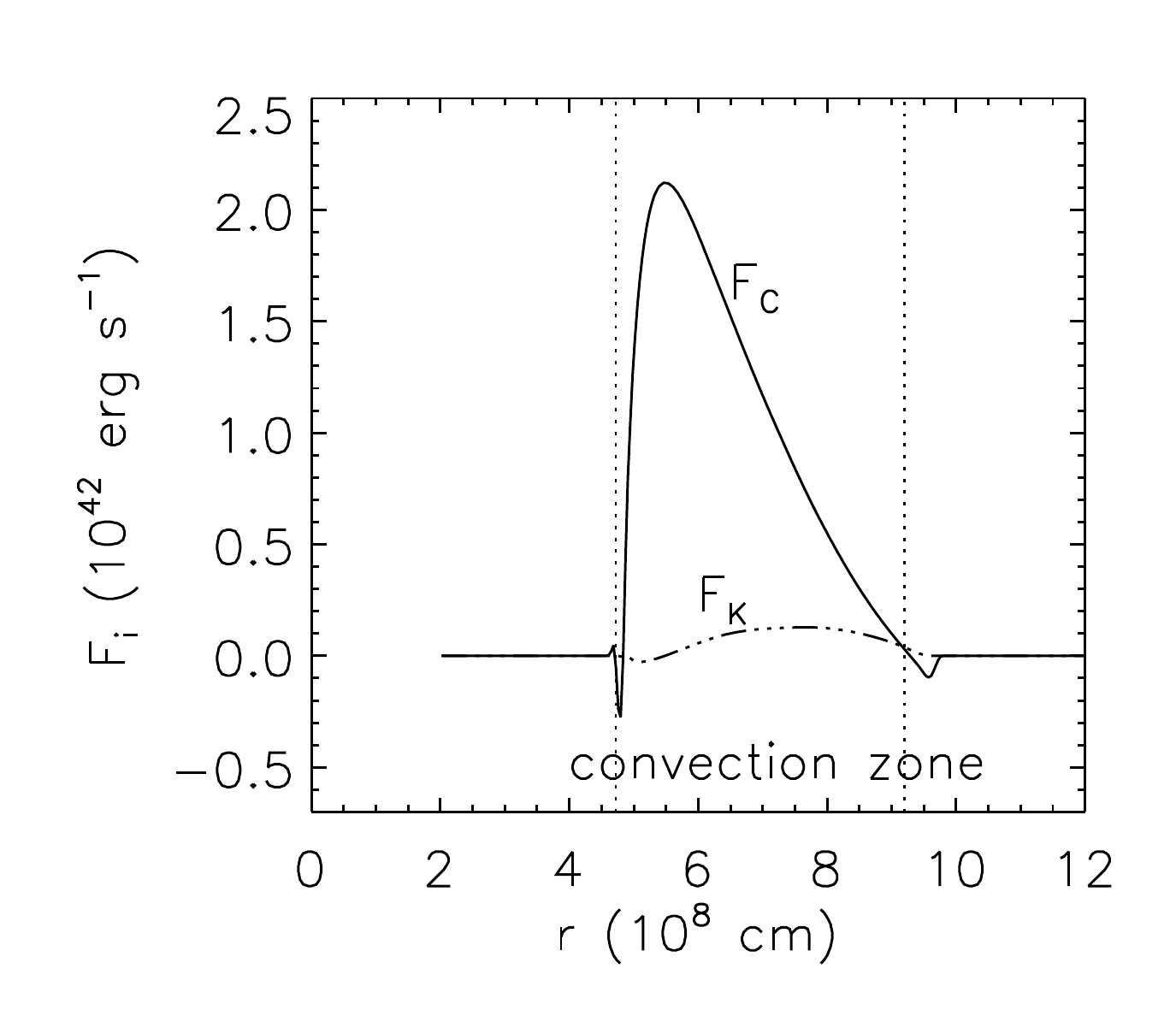} 
\caption{Convective and kinetic energy fluxes ($F_C$ and $F_K$,
  respectively) as a function of radius averaged (from 33\,000\,s to
  53\,000\,s) over about 20 convective turnover timescales for the 3D
  model heflpopI.3d. The dotted vertical lines mark the edges of the
  single convection zone in the initial model M according to the
  Schwarzschild criterion.  }
\label{fig.flx}
\end{figure}

\begin{figure}
\includegraphics[width=0.99\hsize]{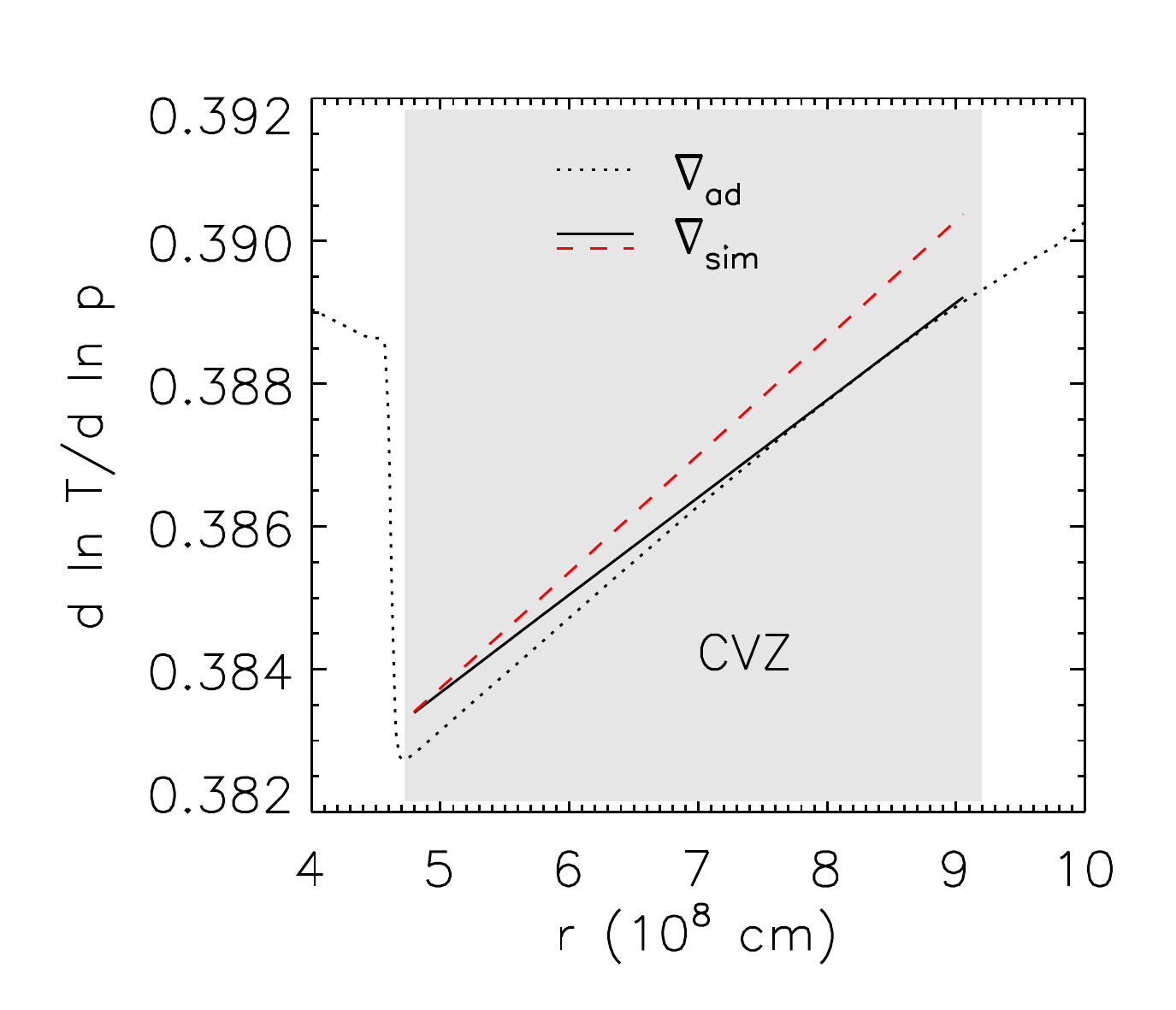} 
\caption{Radial distribution of the adiabatic temperature gradient
  $\nabla_{ad}$ (dotted) obtained using the EOS, and of the
  temperature gradient of the 3D model heflpopI.3d averaged over the
  first 460\,s of its evolution (dashed-red), and over a period of
  13000\,s between 33000\,s and 46000\,s (solid-black), respectively.
  The latter gradients shown are actually linear fits to
  the model data. The gray shaded region marks the single
  convection zone CVZ.  }
\label{fig.tmpgradpop1}
\end{figure}

\section{Results}
\label{sect:tmpevol}

In this section, we first present the characteristics of the
hydrodynamic simulation based on the initial model M, \ie model
heflpopI.3d, which shows a fast growth of the convection zone
(Sect.\,\ref{subs:sflash}). A growth rate of this magnitude likely
leads to a hydrogen injection phase (Sect.\,\ref{subs:sadflash}),
which may resemble the one seen in the initial model SC, whose
hydrodynamic properties are discussed in Sect.\ref{subs:dflash}.

\subsection{Single flash}
\label{subs:sflash}
%
Table\,\ref{modpop1tab} provides some characteristic parameters of our
3D simulation heflpopI.3d based on the initial model M.  After
convection reaches a quasi steady-state in this model, the maximum
temperature rises at the rate of 80\,\Ks, \ie only 20\% slower than
predicted by canonical stellar evolution theory. This corresponds to
an increase of the nuclear energy production rate from $2.4 \times
10^{42} \ergs$ at $t \sim 20\,000\,$s to $5.5 \times 10^{42} \ergs$ at
$t \sim 99\,000\,$s.  Consequently, the maximum convective velocities
rise by 26\% from about $10^6 \cms$ to $1.26 \times 10^6 \cms$.  As
illustrated in Fig.\,\ref{fig.veloc} these velocities match those
predicted by the mixing length theory quite well.  During the first
third of the simulation (up to about $30\,000\,$s) the angle and time
averaged radial velocity in the convective layer exceeds the velocity
predicted for inital model M by the mixing-length theory, $v_{mlt}$,
by about 20\%, while the velocity modulus, $v_{abs}$, is about 30\%
larger than $v_{mlt}$. Towards the end of our simulation the angle and
time averaged modulus of the velocity is about twice as large as
$v_{mlt} $.

\begin{figure}
\includegraphics[width=0.99\hsize]{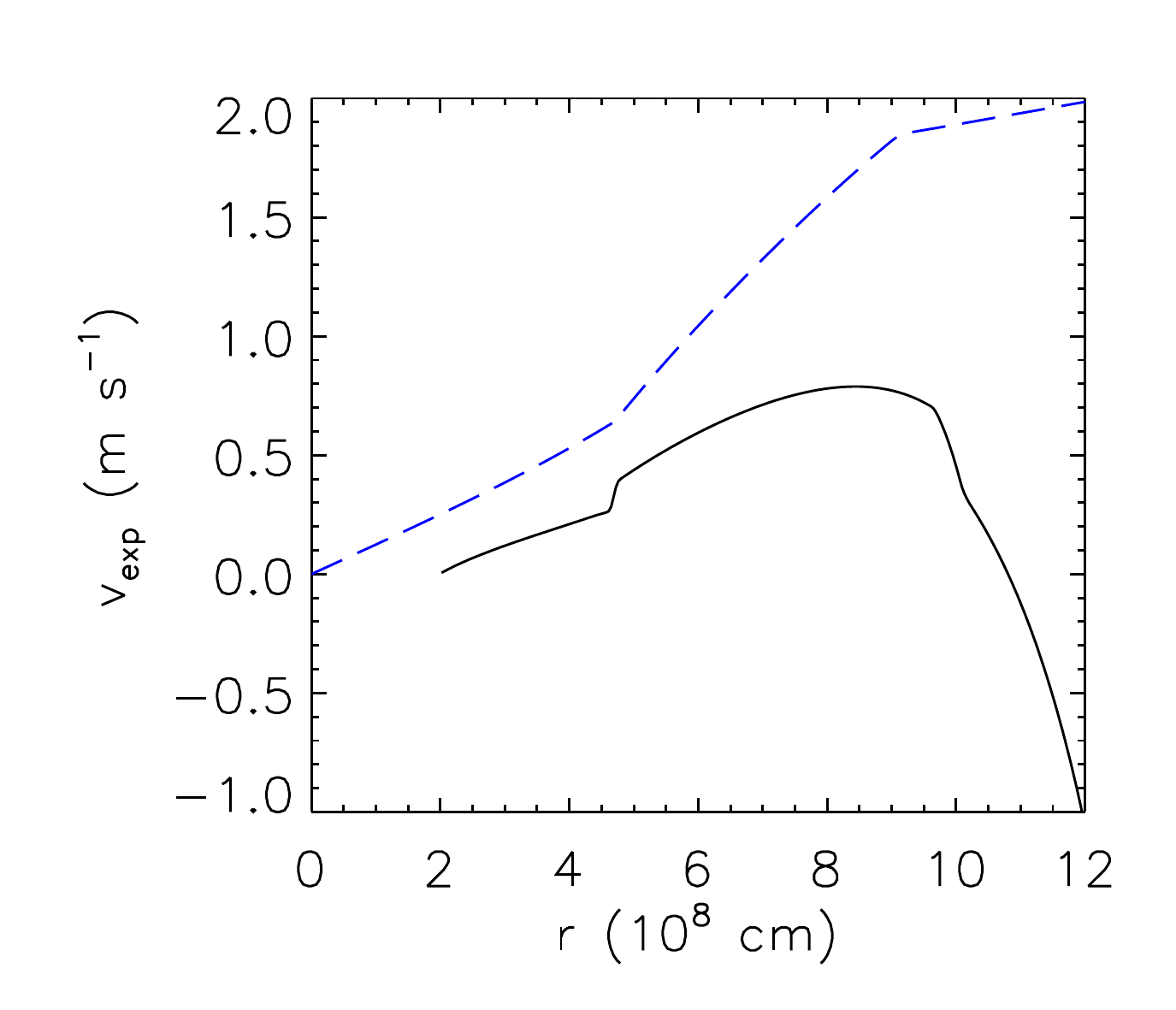} 
\caption{Radial distribution of the expansion velocity, $v_{exp}$, in
  the 3D model heflpop1.3d (solid) compared with the expansion
  velocity predicted by the stellar evolutionary calculations
  (dashed-blue) for the initial model M. }
\label{fig.exppop1}
\end{figure}

Contrary to our previous study \citep{Mocak2009}, we do not find a
sub-adiabatic gradient in the outer part of the convection zone.
\footnote{A sub-adiabatic gradient does not imply that convective
  blobs are cooler than their environment and that consequently
  convection ceases (the latter is only true when assuming
  adiabatically rising blobs). It only means that blobs cool faster
  than their environment. }.
The reason for this difference is probably the increased grid
resolution of our present 3D simulation, which results in less heat
diffusion due to numerical dissipation, and hence a super-adiabatic
temperature gradient similar to the initial one
(Fig.\,\ref{fig.tmpgradpop1})

\begin{figure*}
\includegraphics[height=5.2cm]{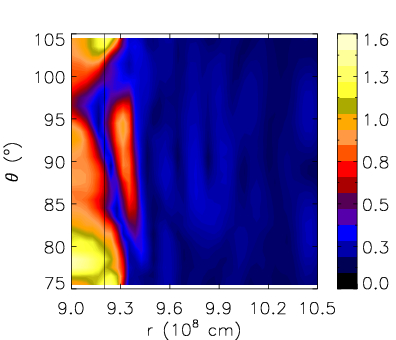} 
\includegraphics[height=5.2cm]{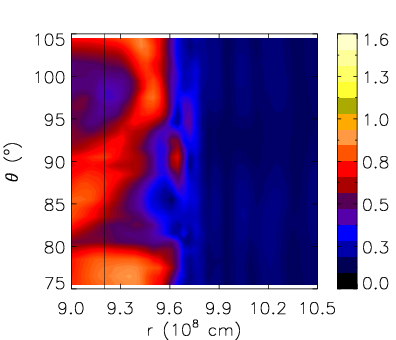} 
\includegraphics[height=5.2cm]{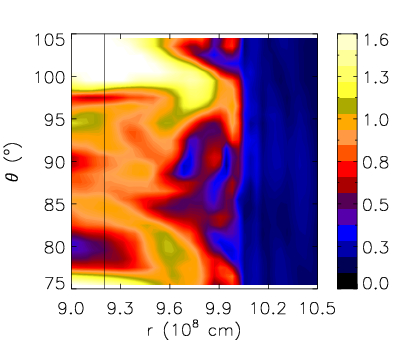} 
\caption{Color maps of the modulus of the velocity (in units of
  $10^{6} \cms$) near the outer boundary of the convection zone for
  the 3D model heflpopI.3d in the meridiononal plane $\phi = 0\dgr$ at
  three different epochs: $t_1 = 3295\,$s (left), $t_2 = 49\,385\,$s
  (middle), and $t_3 = 97\,655\,$s (right).}
\label{fig.svelbndry}
\end{figure*}

\begin{figure*}
\includegraphics[height=7.5cm]{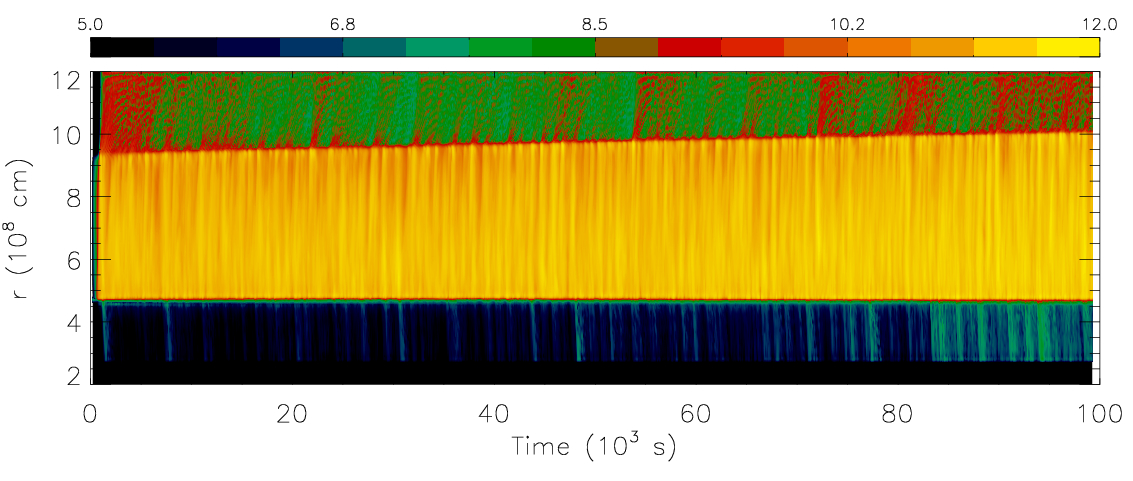} 
\caption{Temporal evolution of radial distribution of the (color
  coded) logarithm of the angular averaged kinetic energy density (in
  \ergg) in the 3D model heflpopI.3d. The growth of the size of the
  convection zone due to the turbulent entrainment, mainly at its
  outer boundary, is clearly visible.}
\label{fig.kened}
\end{figure*}

At a radius $r \sim 5.5\times 10^8\,$cm convection transports almost
90\% of the liberated nuclear energy, \ie $2.1\times 10^{42} \ergs$
(Fig.\,\ref{fig.flx}). The energy flux due to thermal transport is
negligible (Sect.\,\ref{sect:input}).

We observe internal gravity waves
\footnote{In a convectively stable region, any displaced mass element
  is pushed back by the buoyancy force. On its way back to its
  original position, the blob gains momentum and therefore starts to
  oscillate. These oscillations are called internal gravity waves
  \citep{Dalsgaard2003}.}
or g-modes in the convectively stable layers. These g-modes are
strongly instigated only during certain evolutionary phases (g-mode
events) because of the intermittent nature of the convective flow. The
g-mode events are correlated with outbursts in kinetic energy of the
convection zone \citep{MeakinArnett2007}. It appears that the
frequency of g-mode events increases as the flow gains strength
(Fig.\,\ref{fig.kened}).

\begin{figure*}
\includegraphics[width=0.49\hsize]{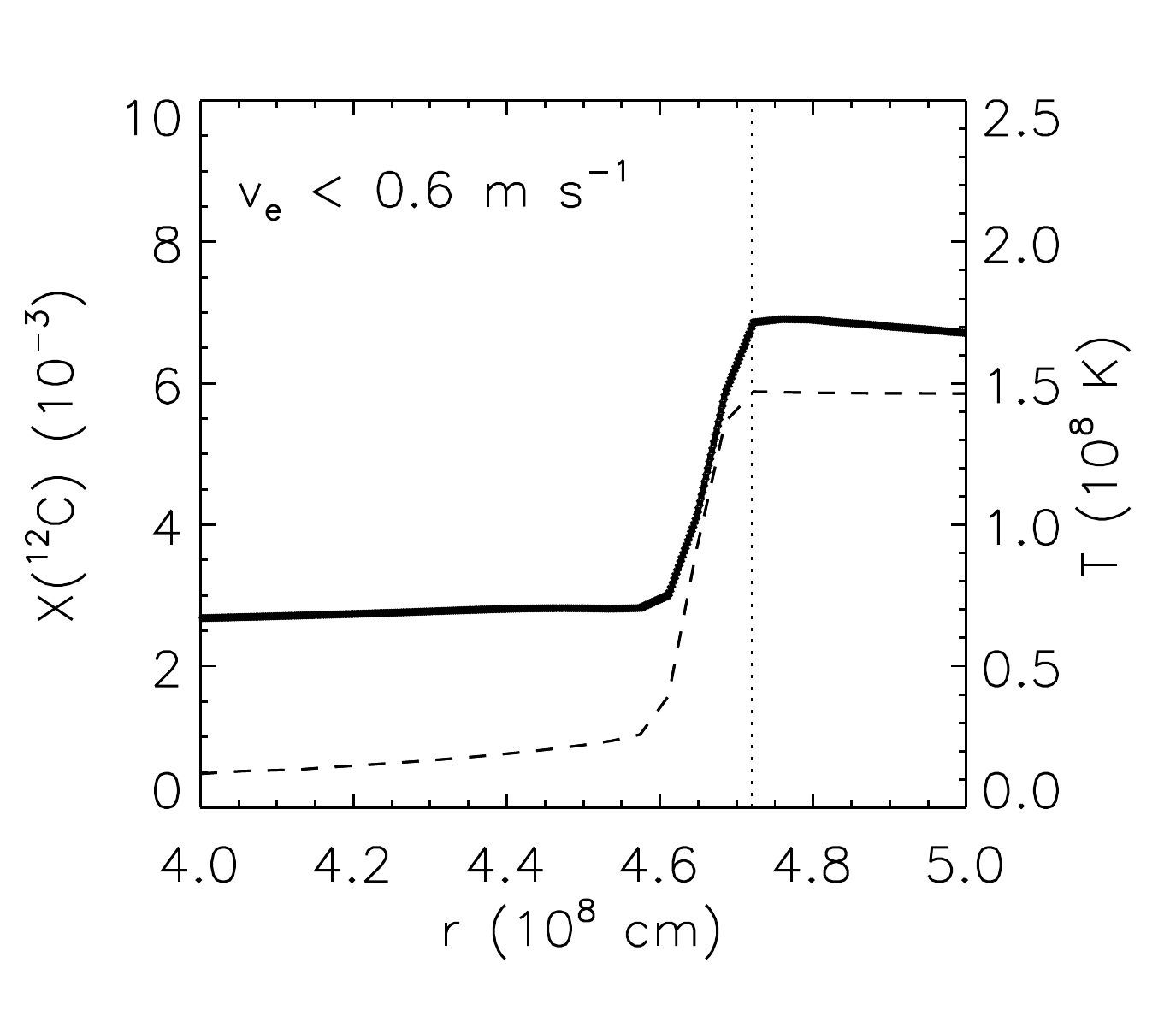}
\includegraphics[width=0.49\hsize]{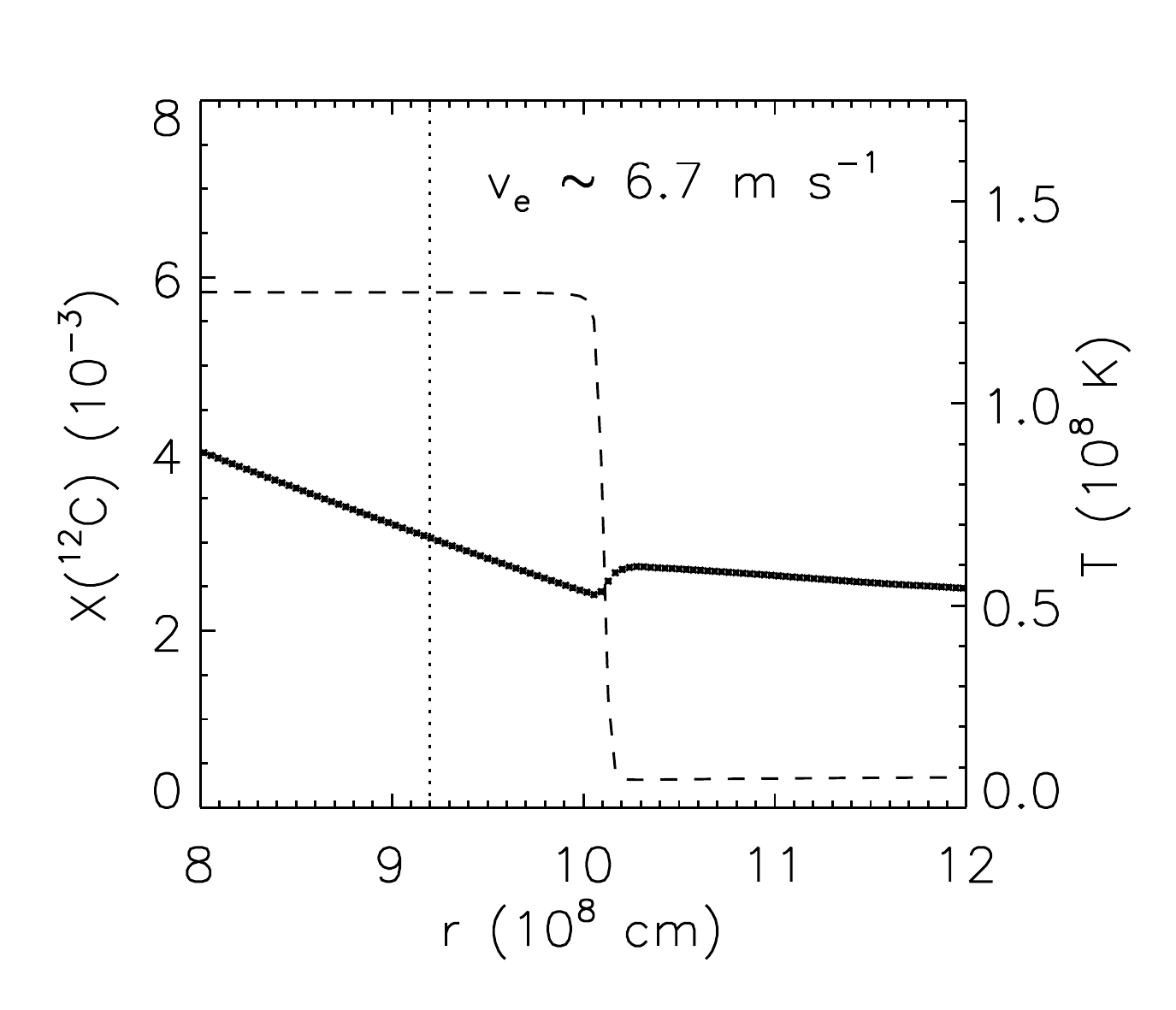} 
\caption{Angular averaged $^{12}$C mass fraction as a function of
  radius near the inner (left) and outer edge (right) of the
  convection zone in the 3D model heflpopI.3d at $t = 100\,000\,$s.
  The thick line gives the corresponding temperature stratification,
  and the vertical dotted lines mark the edges of the convection zone
  at $t = 0\,$s.  The observed entrainment velocities v$_e$ are given 
  in both panels, too. }
\label{fig.cargrowth}
\end{figure*}

We observe a growth of the size of the single convection zone due to
turbulent entrainment at the convective boundary on a dynamic
timescale (Figs.\,\ref{fig.svelbndry} to \ref{fig.cargrowth}), and
estimate the corresponding entrainment speeds by adopting the
prescription of \citet{MeakinArnett2007}. Turbulent entrainment
involves mass entrainment (Fig.\,\ref{fig.svelbndry}) rather than a
diffusion process, which acts to reduce the buoyancy jump at the
convective boundary allowing matter to be mixed further.  The
entrainment velocity or the interface migration velocity, $u_e$, is
given by (see Eq.\,(32) of \citet{MeakinArnett2007}
\begin{equation}
  u_e = \frac{\Delta q}{h N^2} ,
\label{eq:espeed}
\end{equation}
where $\Delta q$ is the buoyancy jump, \ie the variation of the
buoyancy flux $q$ across the convective boundary of thickness $h$, and
$N^2$ the square of the Brunt-V\"ais\"al\"a buoyancy frequency.  The
buoyancy flux is defined as $q = g \rho' v' / \rho$, where $g, \rho$,
and $v$ are the gravitational acceleration, the density, and the
velocity, respectively. The quantities $\rho'$ and $v'$ denote
fluctuations around the corresponding mean quantities $\rho_0$ and
$v_0$, \ie $\rho = \rho_0 + \rho'$ and $v = v_0 + v'$,
respectively. The thickness $h$ of the convective boundary is defined
as the half width of the peak in the mass fraction gradient of
$^{12}$C, which varies rapidly at the boundary.  The entrainment
velocities (averaged over 20 convective turnovers at the end of the
simulation) obtained by this formula are summarized in
Tab.\,\ref{tab:entr} together with those derived from measuring the
position of the convective boundary in our hydrodynamic simulations.

\begin{table} 
\caption{Some quantities characterizing the convective boundaries of
  the 3D model heflpopI.3d : buoyancy jump $\Delta q$ (in $\erggs$),
  square of the Brunt-V\"ais\"al\"a buoyancy frequency $N^2$,
  interface migration velocity calculated according to
  Eq.\,\ref{eq:espeed}, $u_e$, and obtained from our simulation,
  $v_e$, and expansion velocity $v_{exp}$, at the inner and outer
  convective interface of thickness $h$, respectively.  }
\begin{center}
\begin{tabular}{p{0.7cm}|p{0.7cm}p{0.7cm}p{1.1cm}p{0.7cm}p{0.7cm}p{0.9cm}} 
\hline
\hline
pos & $\Delta q$ & $h$ & $N^2$ & $u_e$ & $v_e$ & $v_{exp}$  \\
 & [$10^8$] & [$10^7$cm] & [$\rads$] & [$\mes$] & [$\mes$] & [$\mes$] 
\\
\hline 
inner & 9.7 & 1.0 & 0.92  & 1.1 & $<$0.6 & 0.4  \\
outer & 6.5 & 1.2 & 0.14  & 3.9 & ~~~6.7 & 0.8  \\ 
\hline
\end{tabular} 
\end{center}
\label{tab:entr} 
\end{table}  

The entrainment velocity derived for our models, $v_e$, are calculated
by measuring the radial position of the convective boundaries defined
by the condition X($^{12}$C) $= 2\times 10^{-3}$, as $^{12}$C is much
less abundant outside the convection zone (Fig.\,\ref{fig.cargrowth}).
The expansion velocity (zero in hydrostatic equilibrium) is given by
$v_{exp} = \dot{m}_r / 4 \pi r^2 \rho$, where $\dot{m}_r$ is the
partial time derivative of the mass $m_r$ of a shell of density $\rho$
at a radius $r$.

We find in our models that $v_e$, which includes the expansion
velocity $v_{exp}$, agrees very well with the velocity $u_e$ predicted
by theoretical considerations of the entrainment process (see
Eq.\,\ref{eq:espeed}), despite the crude estimate of the divergence of
the buoyancy flux
\footnote{$\partial_t b = div(q) \sim u_e \partial_r b = u_e N^2$ is a
  rather crude approximation which provides an order of magnitude
  estimate only \citep{MeakinArnett2007}.}
through $\Delta q/h$.  The velocity given by the difference $v_e -
v_{exp}$, which is the actual entrainment speed in our models, differs
from the theoretically estimate $u_e$ by 0.9$\mes$ and 2.0$\mes$ at
the inner and outer convection boundary, respectively.

It seems that turbulent entrainment is a robust process which has been
seen to operate under various conditions in different stars
\citep{Mocak2009, MeakinArnett2007}. This process may be behind the
observed Al/Mg anti-correlation \citep{Shetrone1996a, Shetrone1996b,
  Yong2006}, which could result from an injection of hydrogen into the
helium core and a subsequent dredge-up \citep{Langer1993, Langer1995,
  Langer1997, Fujimoto1999}.

\begin{table*} 
\caption{Some properties of the 2D and 3D hydrodynamic models based on
  initial model SC: wedge size $w$, number of grid points in $r$,
  $\theta$, and $\phi$ direction, estimated Reynolds numbers $R_{e1}$,
  $R_{e2}$ \citep{PorterWoodward1994}, characteristic flow velocities,
  $v_{c1}$ and $v_{c2}$, and typical convective turnover timescales,
  $\tau_{cnv1}$ and $\tau_{cnv2}$, for the bottom and top part of the
  convection zone, respectively.  $t_{max}$ is the maximum evolutionary time. }
\begin{center}
\begin{tabular}{p{2.cm}|p{1.8cm}p{0.5cm}p{1.3cm}p{0.4cm}p{0.4cm}p{0.4cm}
                p{0.4cm}p{1.4cm}p{1.4cm}p{0.8cm}p{0.8cm}p{0.8cm}} 
\hline
\hline
run & grid & $w$ & $\Delta r$ & $\Delta\theta$ & $\Delta\phi$ & $R_{e1}$ 
& $R_{e2}$ &  $v_{c1}$ &  $v_{c2}$  & $\tau_{cnv1}$ & $\tau_{cnv2}$ & $t_{max}$ \\
$\mbox{[name]}$ & N$_r \times$N$_\theta \times$N$_\phi$ & [$\dgr$] & 
[$10^{6}$ cm] 
& [$\dgr$] & [$\dgr$]  & & & [$10^{6}\,\cms$] & [$10^{6}\,\cms$] & [~s~] 
& [~s~] & [~s~]
\\
\hline
heflpopIII.2d.1 &  $720 \times 50$           &  50 & 7.6 & 1 & - &
  $10^3$ & $10^2$ & 1.3       & 1.0       & 2900 & 2400 & 1400 \\ 
heflpopIII.2d.2 & $1500 \times 120$          & 120 & 3.7 & 1 & - & 
  $10^5$ & $10^3$ & 1.7  & 1.0  & 2200 & 2400 & 6400 \\ 
heflpopIII.3d   &  $720 \times 50 \times 50$ &  50 & 7.6 & 1 & 1 & 
  $10^2$ & $10^2$ & 0.4       & 0.3       & 9500 & 8000 & 1400 \\   
\hline
\end{tabular} 
\end{center}
\label{modpop3tab} 
\end{table*}  

\begin{figure*}
\includegraphics[height=9.0cm]{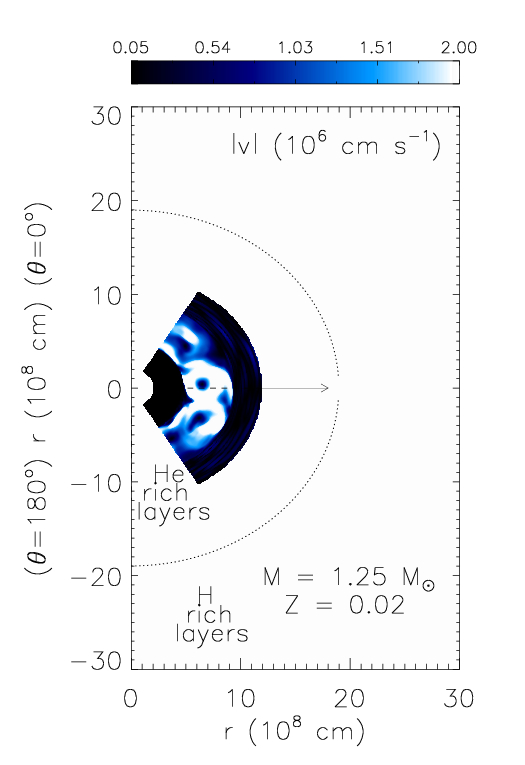} 
\includegraphics[height=9.0cm]{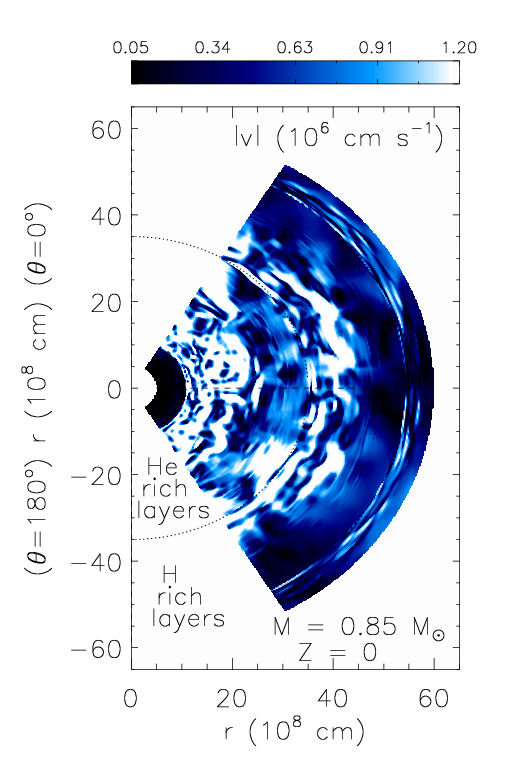}
\includegraphics[height=9.0cm]{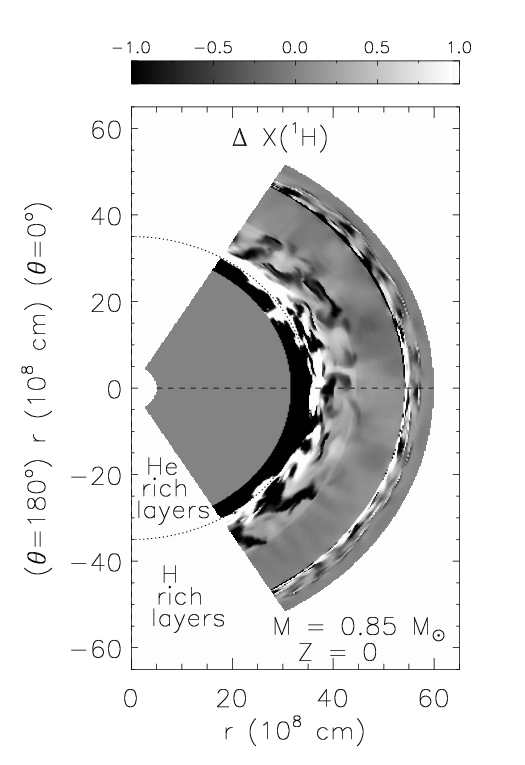}
\caption{Snapshots of the spatial distribution of the velocity modulus
  $|\mbox{v}|$ (in units of $10^{6} \cms$) of a typical 2D
  hydrodynamic model with a single convection zone (left), and for the
  2D hydrodynamic model heflpopIII.2d.2 at 1250\,s (middle). The right
  hand panel shows the double convection zone highlighted by using the
  hydrogen contrast $\Delta X(^{1}\mbox{H}) = \mbox{100}\times
  (X(^{1}\mbox{H}) - \langle X(^{1}\mbox{H}) \rangle_{\theta}) /
  \langle X(^{1}\mbox{H}) \rangle_{\theta}$ at the same time, where
  $\langle \rangle_{\theta}$ denotes a horizontal average at a given
  radius. The arrow indicates the growth of the single convection
  zone, while the dotted line represents the border between the helium
  and hydrogen rich layers.  }
\label{fig.snap2d}
\end{figure*}

\subsection{From the single to the dual flash}
\label{subs:sadflash}
%
If the radial position of the innermost edge of the hydrogen-rich
layers was fixed at its initial value in model M at $r = 1.9\times
10^9\,$cm (no expansion), the outer convection boundary would reach
the hydrogen-rich layer due to the turbulent entrainment within only
17 days, and the helium core would experience a dual core flash (DCF)
known from low-mass Pop III stars.

However, the hydrogen layer will initially expand outwards at a faster
rate than the outer convective boundary (Fig.\,\ref{fig.exppop1}).
This delays the expected onset of the hydrogen injection a little, as
the outer convection boundary has to catch up with the hydrogen layer
expanding away from the HeCZ.  As the expansion velocities of our
hydrodynamic models are biased by the imposed reflective
boundaries in radial direction, we did not use these values here. To
get an estimate for the onset of the hydrogen injection we instead
used the expansion speed of the helium core as predicted by stellar
evolutionary calculations (Tab.\,\ref{tab:entr}), and find that the
injection of hydrogen into the helium core should take place within 23
days.

\begin{figure}
\includegraphics[width=0.99\hsize]{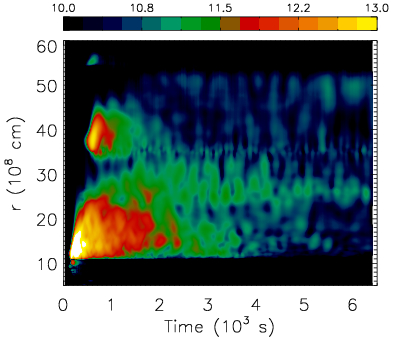} 
\includegraphics[width=0.99\hsize]{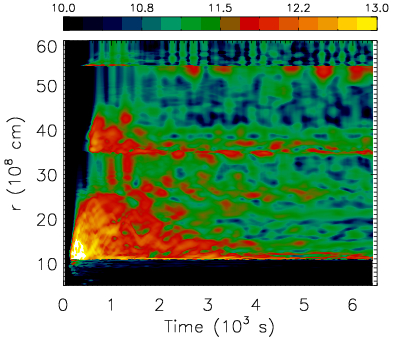} 
\caption{Temporal evolution of the radial distribution of the (color
  coded) logarithm of the angular averaged radial ($v_r^2/2$; upper
  panel) and angular $v_\theta^2/2$; lower panel) component of the
  specific kinetic energy (in \ergg) of the 2D model heflpopIII.2d.2.
}
\label{fig.kindnstdec}
\end{figure}

\begin{figure}
\includegraphics[width=0.99\hsize]{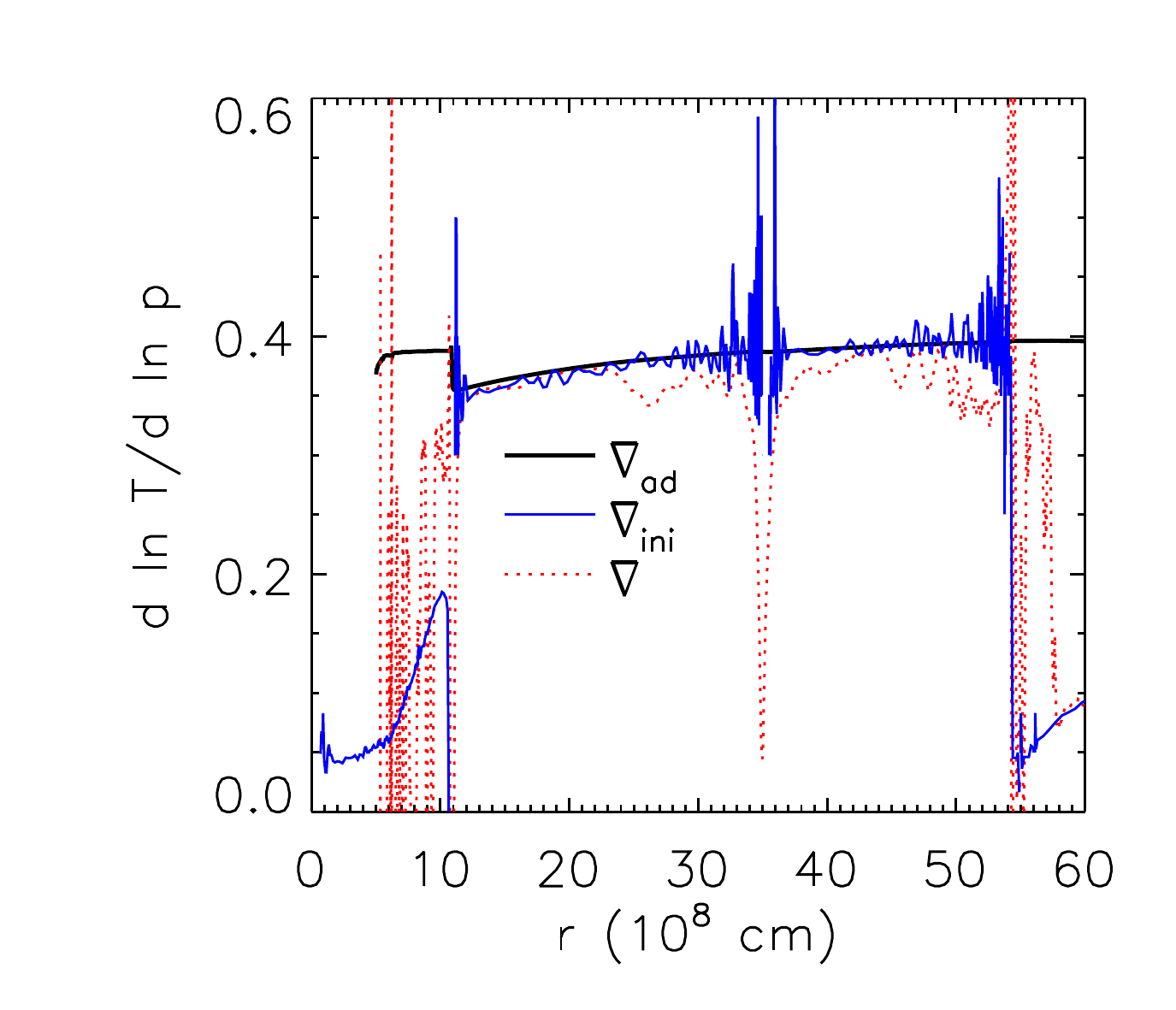} 
\caption{Adiabatic temperature gradient $\nabla_{ad}$ (solid-black),
  temperature gradient of the 2D model heflpopIII.2d.2 averaged from
  1000\,s to 3200\,s (dashed-red), and temperature gradient of the
  initial model SC (solid-blue), respectively, as a function of
  radius. }
\label{fig.tmpgradpop3}
\end{figure}

This is still a very short time, and even if our estimate for the time
until the onset of the hydrogen injection was wrong by 5 orders of
magnitude, the injection will occur before the first subsequent mini
helium flash
\footnote{When the helium burning shell of the first major helium
  flash becomes too narrow, it is not able to lift the overlying mass
  layers. It expands slowly, \ie $\delta \rho / \rho < 0$, but remains
  almost in hydrostatic equilibrium $\delta P $/$ P \sim 0$, which in
  turn leads to a rise of its temperature $\delta T $/$ T > 0$
  \citep{KipWeigert1990}. Hence, helium is re-ignited, but less
  violently than in the first main helium flash.  In general, one
  refers to this event as a thermal pulse, but we prefer to call it a
  mini helium flash.  This process repeats itself several times until
  the star settles on the horizontal branch. }
takes place, which is supposed to occur in roughly $10^5$~years.

Figure\,\ref{fig.snap2d} shows snapshots from 2D hydrodynamic
simulations \citep{Mocak2009} of a (single) core helium flash based on
initial model M (left panel), and of a dual core (helium + hydrogen)
flash based on initial model SC (middle and right panels).  The figure
suggests that that even if the core helium flash starts out with a
single convection zone in a low-mass Pop I star, this convection zone
may evolve due to its growth by turbulent entrainment (indicated by
the arrow in the left panel of the figure) into a double convection
zone like that found in models of Pop III stars.  Although the upper
boundary of the single convection zone is still far away from the
hydrogen-rich layers at the end of the simulation, it should get there
eventually. We find no reason in any of our 2D or 3D simulations including
the new model heflpopI.3d why turbulent entrainment should cease 
before the outer convective boundary reaches the edge of
the helium core. Actually, as the maximum temperature of the helium
core grows and the convective flow becomes faster, entrainment will
eventually speed up \citep{Mocak2009}.

On the other hand, subsequent mini helium flashes are unlikely to occur,
because we estimate that at the observed entrainment velocity the inner
convective boundary will reach the center of the star in $\gtrsim$ 90
days.  Note in this respect that the fast entrainment speed of the
inner convective boundary derived from our previous 2D models
\citep{Mocak2008} is not confirmed by the new 3D model. The
entrainment rate is actually slower in the new 3D model, which was
expected \citep{Mocak2009}.

The turbulent entrainment at the inner convection boundary heats the
layers beneath at a rate $\delta T/\delta t \sim 630$\Ks, \ie it is
quite efficient in lifting the electron degenaracy in the matter below
the convection zone.

According to the above discussion we propose a somewhat speculative
scenario for the core helium flash in low-mass Pop I stars. These
stars ignite helium burning under degenerate conditions and develop a
single convection zone, which at some point extends due to turbulent
enrainment up to their hydrogen-rich surface layers. The convection
zone eventually penetrates these layers, and dredges down hydrogen
into the helium core. This ignites a secondary flash driven by CNO
burning, which together with triple-$\alpha$ burning and inwards
turbulent entrainment leads eventually to the lifting of the core's
degeneracy, \ie the star will settle down on the horizontal branch.

\subsection{Dual flash}
\label{subs:dflash}
%
We have performed three simulations of the core helium flash based on
the Pop\,III initial model SC which possesses two convection zones
sustained by helium and CNO burning, respectively. Some characteristic
properties of these dual flash simulations are summarized in
Table\,\ref{modpop3tab}.

\subsubsection{Model heflpopIII.2d.2}
\label{subsub:sim2d2}
\begin{figure*}
\includegraphics[width=0.49\hsize]{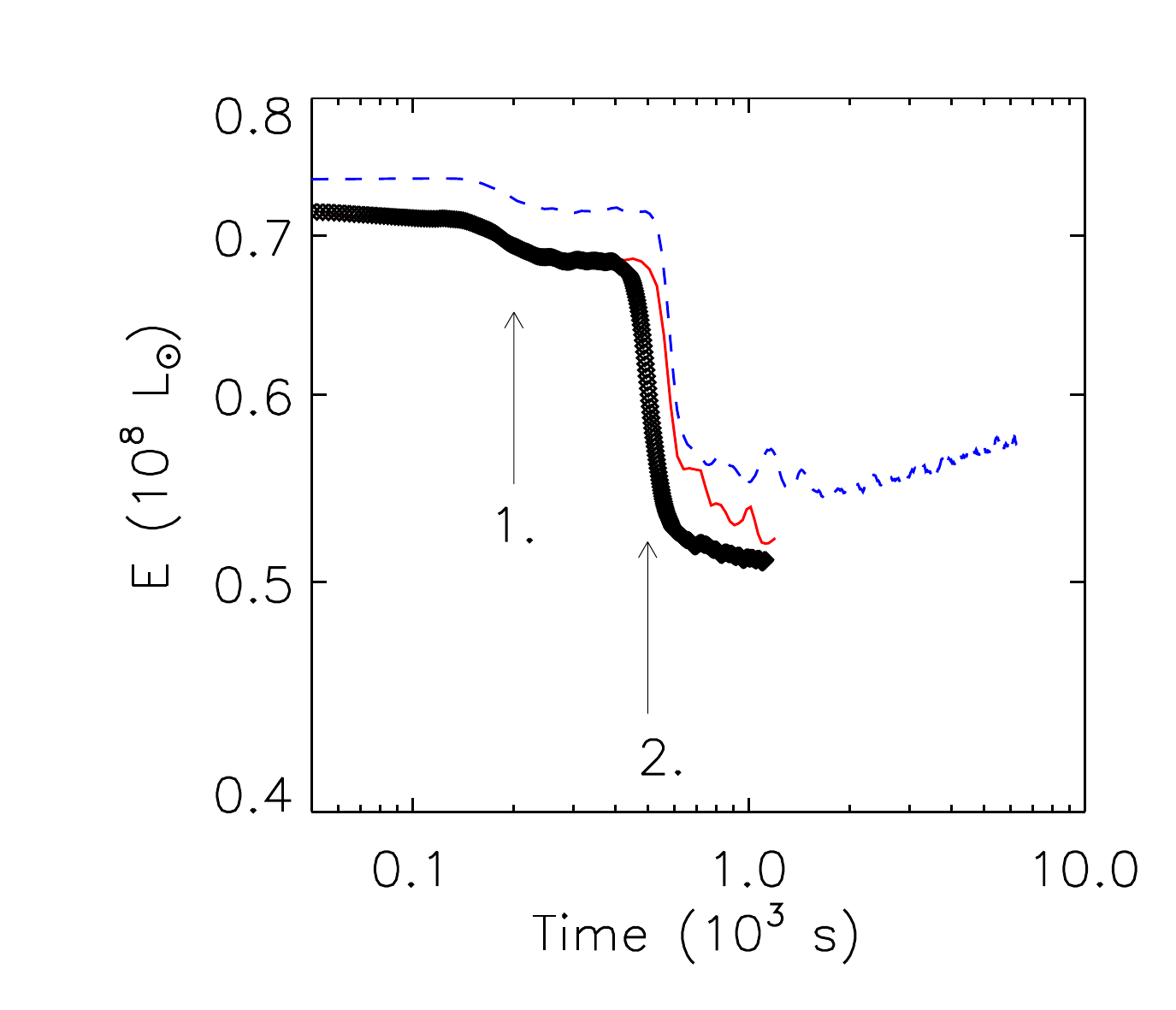}
\includegraphics[width=0.49\hsize]{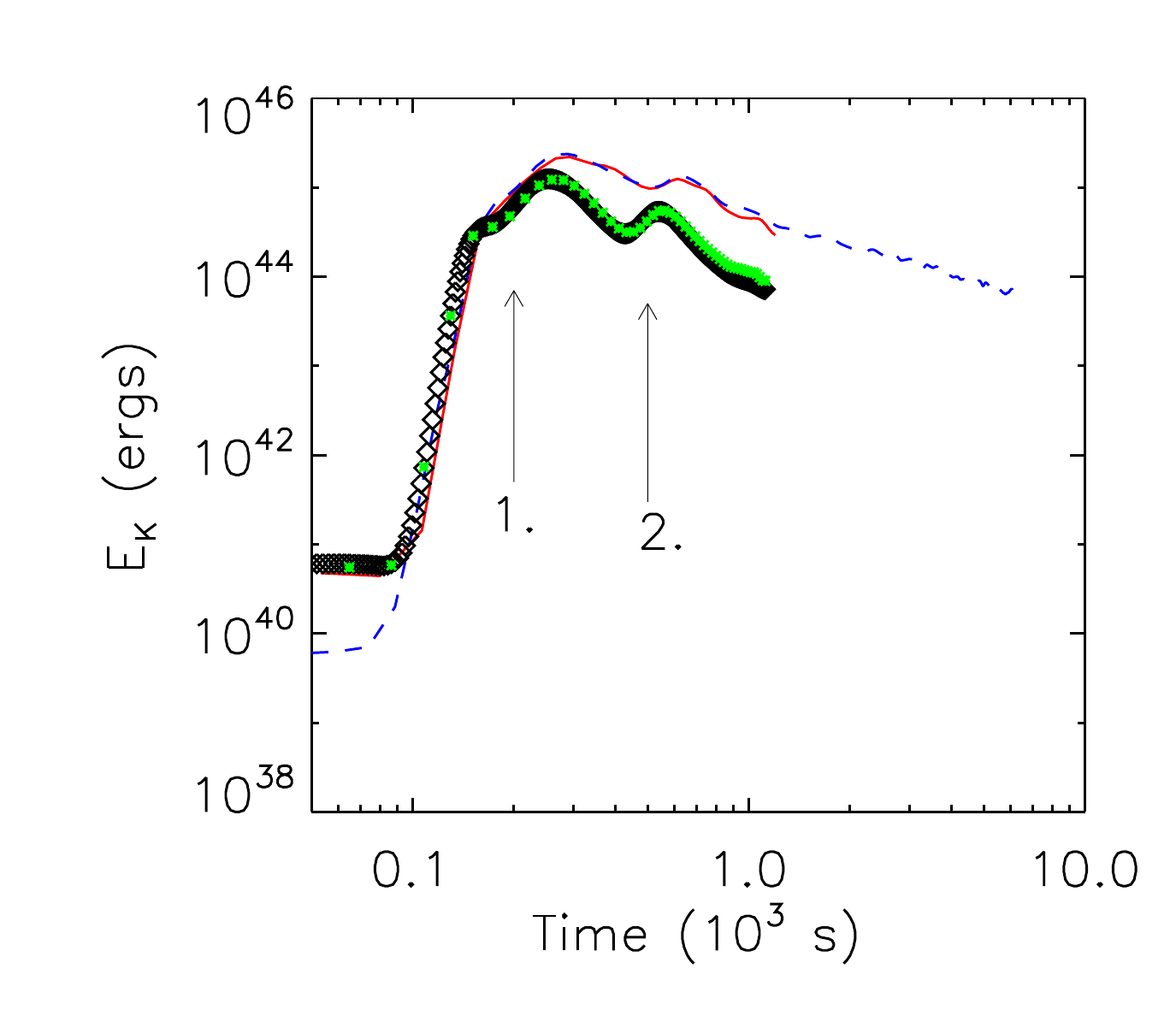} 
\caption{ {\it{Left panel:}} Temporal evolution of the nuclear energy
  production rate $E$ (left) in units of the
  solar luminosity $L_{\odot}$ and the kinetic energy $E_K$ (right) for
  the models heflpopIII.3d (diamonds-black), heflpopIII.2d.1 (solid-red),
  and heflpopIII.2d.2 (dashed-blue), respectively. The green stars 
  give the behavior of a
  3D model with the same properties as model heflpopIII.3d, but with
  nuclear burning switched off. The vertical arrows, labeled 1. and
  2., mark the onset of convection in the lower and upper convection
  zone of the double convection structure, respectively.}
\label{fig.tmpevol}
\end{figure*}

Despite a nuclear energy production rate due to CNO burning in the
outer convection zone ($\dot{\epsilon}_{max} \sim 1.4 \times 10^{10}
\erggs$, which is roughly eight times higher than that due to
triple-$\alpha$ burning in the inner zone $\dot{\epsilon}_{max} \sim
1.7 \times 10^{9} \erggs$), convective motions first appear within the
inner convection zone after about 200\,s.  The onset of convection in
the outer zone is delayed until about 500\,s
(Fig.\,\ref{fig.kindnstdec}, Fig.\,\ref{fig.tmpevol}).  After some
time the model relaxes into a quasi steady-state, where the r.m.s
values of the angular velocity are comparable or even larger than
those of the radial component (Fig.\,\ref{fig.kindnstdec}). This
property of the velocity field implies the presence of g-modes or
internal gravity waves \citep{Asida2000}, which is a surprising fact.
G-modes should not exist in the convection zone according to the
canonical theory, as any density perturbation in a convectively
unstable zone will depart its place of origin exponentially fast
\citep{KipWeigert1990} until the flow becomes convectively stable.

When convection begins to operate in both zones, the total energy
production rate temporarily drops by $\sim$ 20\%, but continues to
rise at a rate of $2.8\times 10^{36}$\ergss \, after $t \approx
2000\,$s. Nevertheless, the convective flow decays fast in both
convection zones -- at a rate of $4\times 10^{40}$\ergs
(Fig.\,\ref{fig.tmpevol}), probably because the initial conditions
represented by the stabilized initial model are too different from
those of the original stellar model.  They disfavor convection since
the stabilized model has a slightly lower temperature gradient than
the original one (Fig.\,\ref{fig.inimt},
Fig.\,\ref{fig.tmpgradpop3}). However, we note that stabilization of
the initial model was essential to prevent it from strongly deviating
from hydrostatic equilibrium.

Shortly after convection is triggered in both the outer and the inner
zone this double convection structure vanishes, and after about
2000\,s there is no evidence left for two separate convection zones
(Fig.\,\ref{fig.kindnstdec}).

Due to the relatively short temporal coverage of the evolution and due
to the decaying convective flow, we did not analyze the energy fluxes
and turbulent entrainment of this model. Penetrating plumes do not
exist in the convection zone 
(initially determined by the Schwarzschild criterion), as
it is dominated rather by g-modes. Hence, firm conclusions are
difficult to derive, but we plan to address this issue elsewhere. We
are now going to introduce some basic characteristic of internal
gravity waves or g-modes that are required to draw further
conclusions.

\paragraph{G-modes} 
%
In a convectively stable region, any displaced mass element or blob
(density perturbation) is pushed back by the buoyancy force. On its
way back to its original position, the blob gains momentum and
therefore starts to oscillate around its original position. Assuming,
the element is displaced by a distance $\Delta r$, has an excess
density $\Delta \rho$, and is in pressure equilibrium with the
surrounding gas ($\Delta$P = 0), one can derive an equation for the
acceleration of the element:
\begin{eqnarray}
  \frac{\partial^2 (\Delta r)}{\partial t^2} =  \frac{g \delta}{H_p}
    \left[\nabla_{\mbox{ele}} - \nabla_{\mbox{surr}} 
                              + \frac{\varphi}{\delta} \nabla_{\mu} 
    \right] \Delta r 
\label{eq.osceq}
\end{eqnarray}
where $g$ is the gravitational acceleration, $H_p = |p /\partial_r p|$
is the pressure scale height, $\nabla \equiv d \ln T / d \ln P$,
$\nabla_\mu \equiv d \ln \mu / d \ln P$, $\delta \equiv \partial \ln
\rho / \partial \ln T$, and $\varphi \equiv \partial \ln \rho /
\partial \ln \mu$.  

\begin{figure}
\includegraphics[width=0.9\hsize]{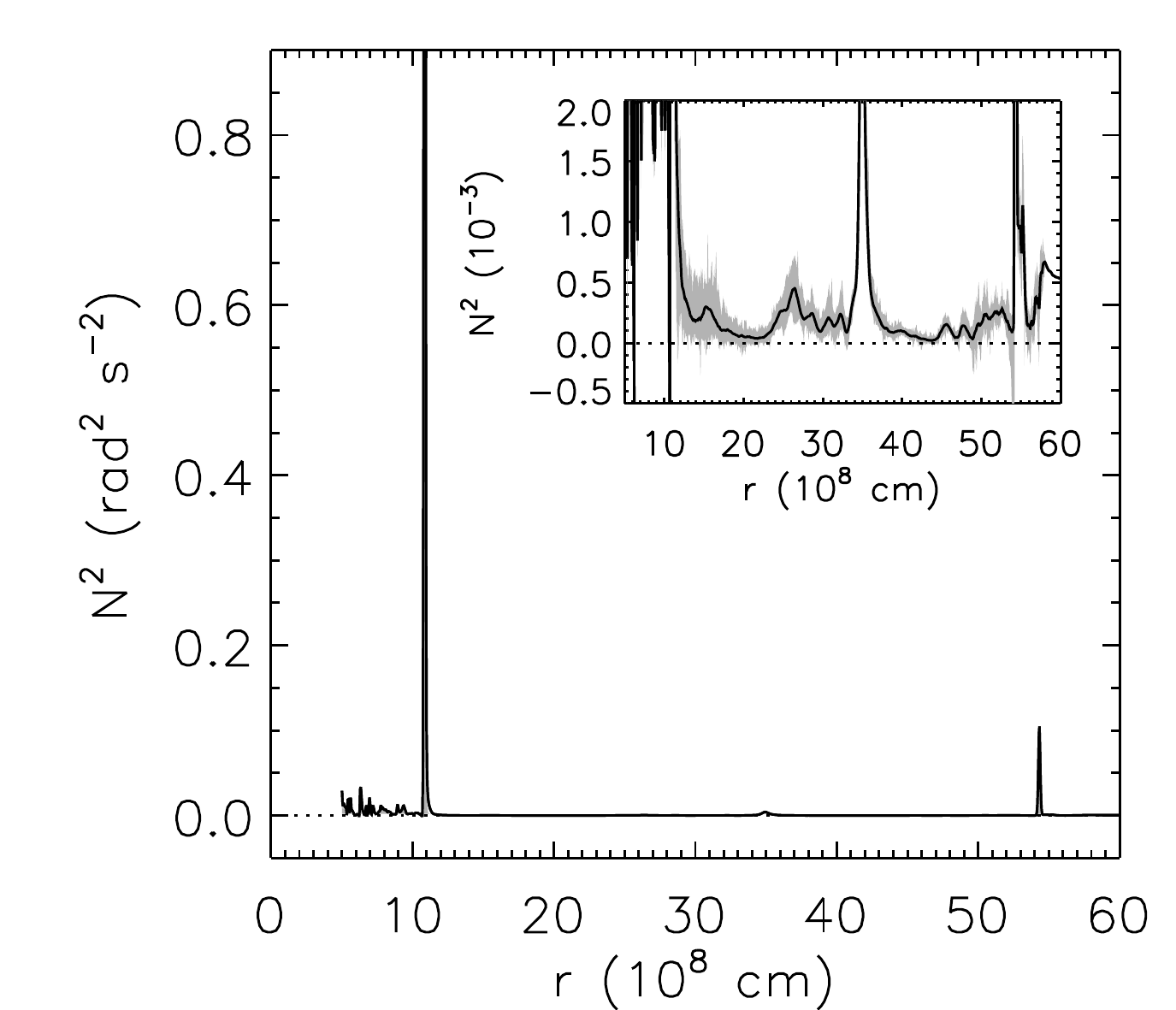} 
\includegraphics[width=0.9\hsize]{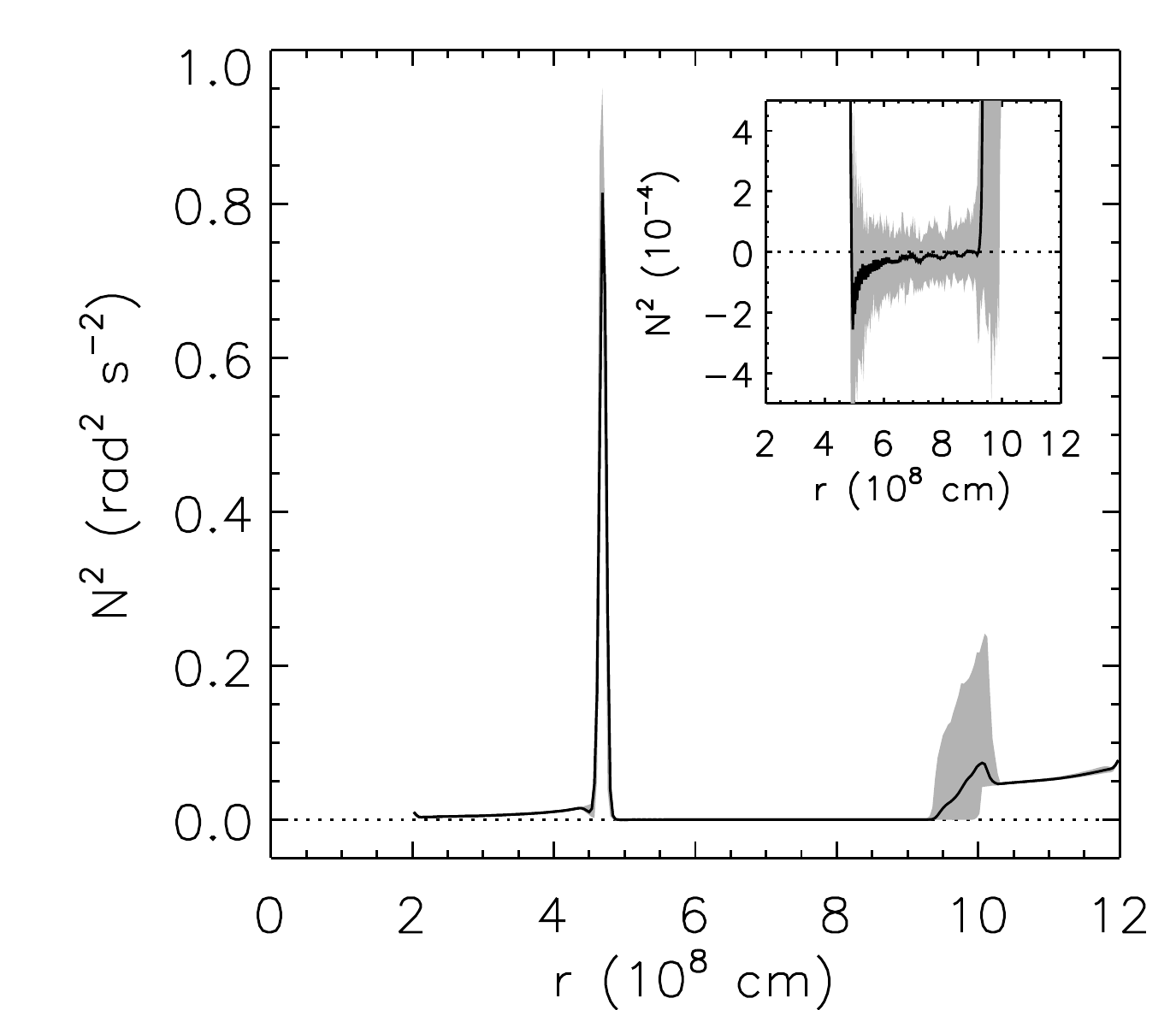} 
\caption{Radial distributions of the (square of the)
  Brunt-V\"ais\"al\"a buoyancy frequency $N^2$ in the 2D model
  heflpopIII.2d.2 averaged between 1480\,s and 6000\,s (top), and in
  the 3D model heflpop1.3d averaged between 6600\,s and 100\,000\,
  (bottom), respectively.  The angular and temporal variation of $N^2$
  at a given radius are indicated by the gray shaded region.  The
  inserts show a zoom of the region around $N^2 = 0$ to enlarge the
  variations of $N^2$ in the convection zone which are $\la
  10^{-3}$. }
\label{fig.brunt}
\end{figure}

\begin{figure}
\includegraphics[width=0.99\hsize]{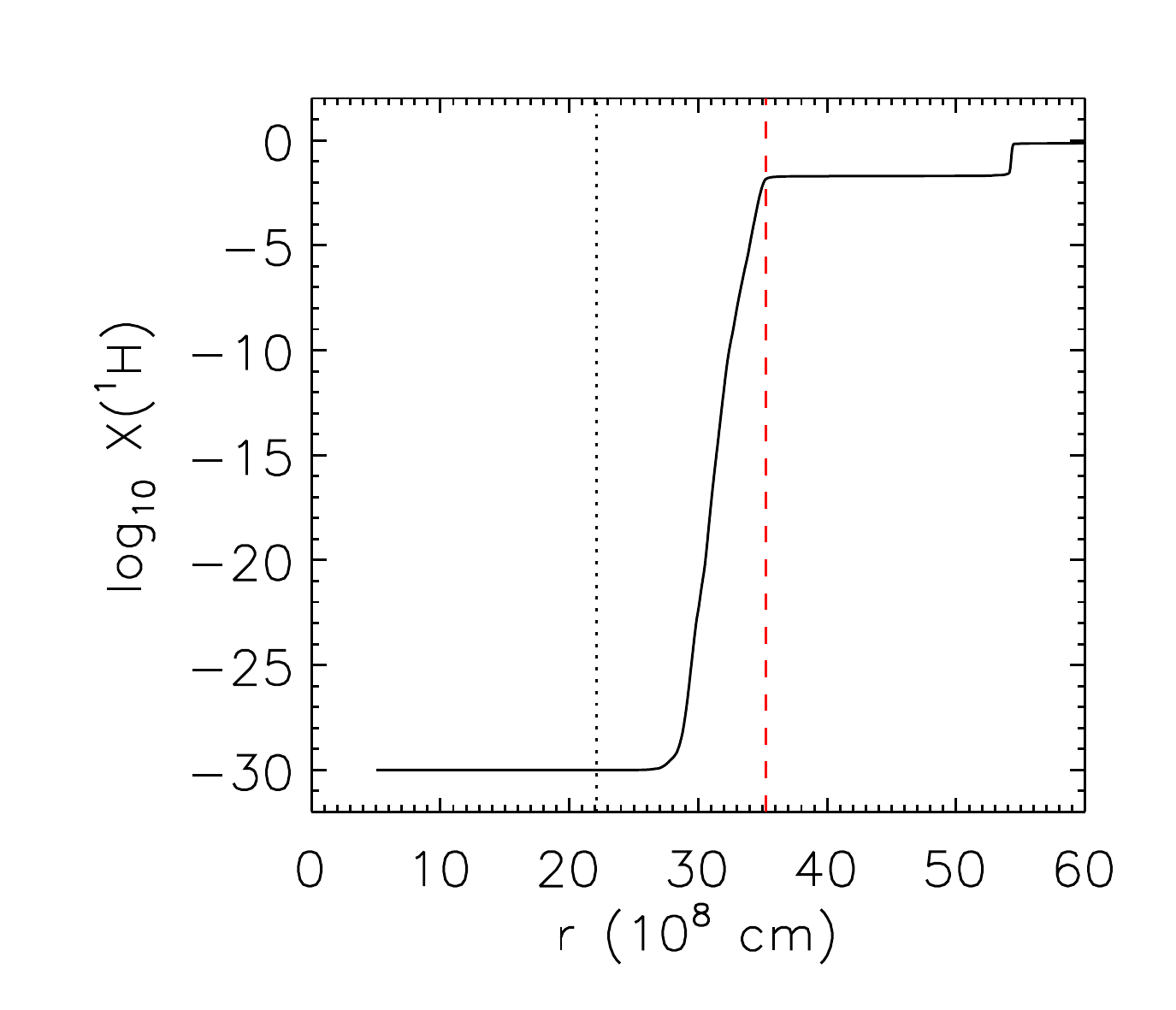} 
\caption{Hydrogen mass fraction as a function of radius for the 2D
  model heflpopIII.2d.2 at $t = 6405\,$s. The vertical lines mark the
  initial border between hydrogen and helium rich layers (dashed-red),
  and the layer (dotted) where the timescale for proton capture on
  $^{12}$C equals 10$^2\,$s ($T \sim 1\times 10^8\,$K).  }
\label{fig.hyddist}
\end{figure}

Let us assume now that the element, after an initial displacement
$\Delta r_0$, moves adiabatically ($\nabla_{\mbox{ele}} =
\nabla_{\mbox{ad}}$) through a convectively stable layer. The element
is accelerated back towards its equilibrium position and starts to
oscillate around this position according to the solution of
Eq.\,(\ref{eq.osceq}):
\begin{eqnarray}
  \Delta r = \Delta r_0 ~e^{i~ N ~t} \, ,
\end{eqnarray}
where the (square of the) Brunt-V\"ais\"al\"a frequency is given by
\begin{eqnarray}
  N^{2} = \frac{g \delta}{H_P} 
          \left( \nabla_{\mbox{ad}} - \nabla_{\mbox{surr}} + 
                 \frac{\varphi}{\delta} \nabla_{\mu} \right) 
\label{eq.n2}
\end{eqnarray}
In a convectively unstable region (assuming $\nabla_{\mu} = 0$),
Eq.(\ref{eq.n2}) implies $N^2 < 0$, \ie $N$ is imaginary.  Thus, the
displaced element moves exponentially away from its initial position,
instead of oscillating around it, as it is the case for  
g-modes or internal gravity waves and $N^2 > 0$.

G-modes appear in layers of gas stratified under gravity and are
spatial oscillatory displacements of density perturbations.  The
dispersion relation of such density displacements, assumed to vary as
$\exp[i({\bf{k}} \cdot {\bf{r}} - \omega t)]$, reads
\citep{Dalsgaard2003}
\begin{equation}
  \omega^2 = \frac{N^2}{1+k^2_r/k^2_h} \, ,
\label{eq.disp}
\end{equation}
where $\omega$ is the temporal frequency of the density displacements,
and k$_r$ and k$_h$ are the radial and horizontal components of the
wave number {\bf{k}}, respectively. The dispersion relation tells us
that any density perturbation must (under influence of buoyancy
forces) displace matter horizontally to propagate vertically.  This
will give rise to matter motion resembling horizontal fingers as
$\omega$ approaches the Brunt-V\"ais\"al\"a frequency for large k$_h$
($k^2_r/k^2_h \rightarrow 0$ and $\omega^2 \rightarrow N^2$).

\begin{figure*}
\includegraphics[height=9.0cm]{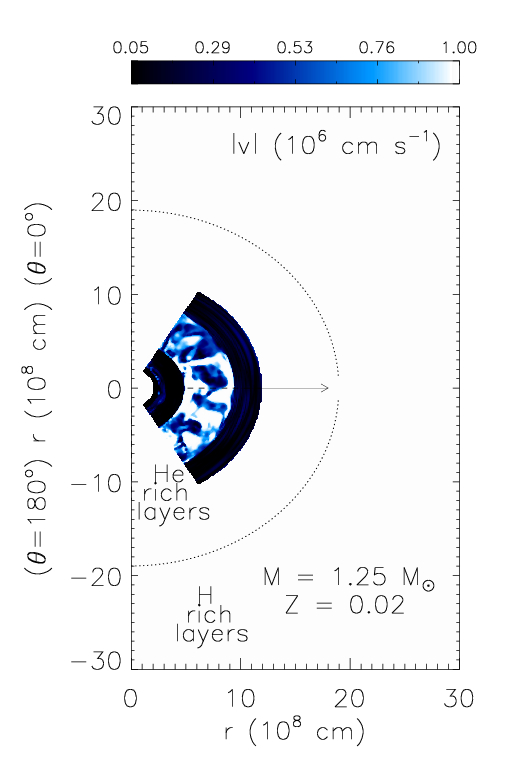} 
\includegraphics[height=9.0cm]{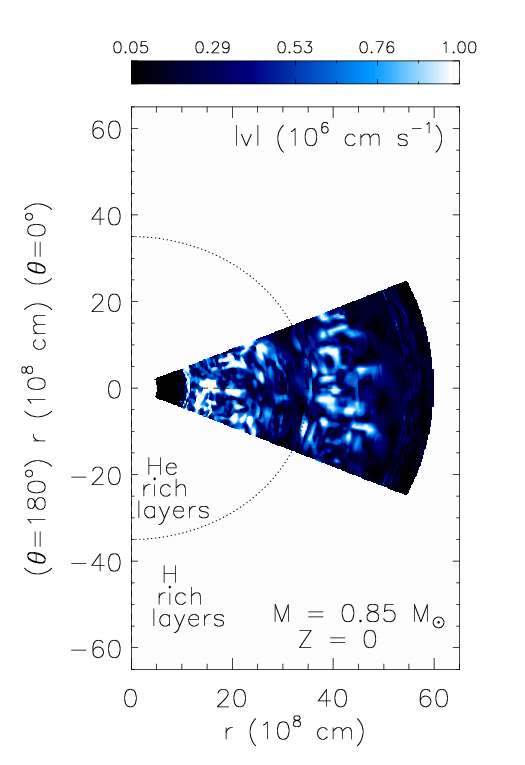}
\includegraphics[height=9.0cm]{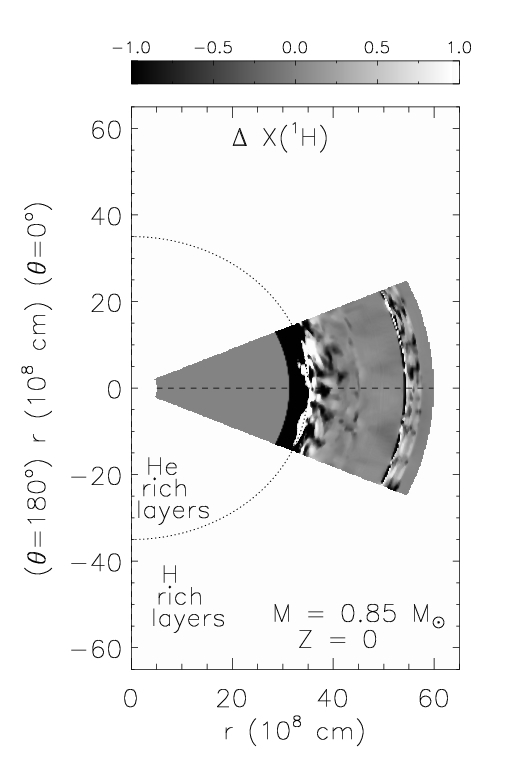}
\caption{Snapshots of the spatial distribution of the velocity modulus
  $|\mbox{v}|$ (in units of $10^{6} \cms$) for a typical 3D
  hydrodynamic model with a single convection zone (left), and for
  model heflpopIII.3d at $t = 1300\,$s with its double convection zone
  (middle) in the meridional plane $\phi = 0\dgr$. The right panel
  shows the hydrogen convection zone highlighted by using the hydrogen
  contrast $\Delta X(^{1}\mbox{H}) = \mbox{100} \times
  (X(^{1}\mbox{H}) - \langle X(^{1}\mbox{H}) \rangle_{\theta}) /
  \langle X(^{1}\mbox{H}) \rangle_{\theta}$, where $\langle
  \rangle_{\theta}$ denotes a horizontal average at a given
  radius. The arrow indicates the growth of the single convection
  zone, while the dotted line represents the border between the helium
  and hydrogen rich layers .}
\label{fig.snap3d}
\end{figure*}

We find such horizontal structures in our models
(Fig.\,\ref{fig.snap2d}) visible mainly in the radiative layer of the
splitted convection zone (Fig.\,\ref{fig.snap2d}). By decomposition of
the specific kinetic energy density of the model into the radial
(v$_r^2$/2) and horizontal (v$_\theta^2$/2) component, we also find
that the horizontal displacements are characterized by higher values
of the kinetic energy density compared to the corresponding values of
the vertical displacements already 2000\,s after the start of the
simulation (Fig.\,\ref{fig.kindnstdec}). Additionally, $N^{2}$ is
mostly positive in these models (Fig.\,\ref{fig.brunt}), indicating
convective stability throughout the double convection zone. This
proves the existence of internal gravity waves in the decaying double
convection zone. The situation is different in our 3D model
heflpop1.3d, where $N^{2}$ is (on average) small and negative
everywhere in the single convection zone (Fig.\,\ref{fig.brunt}).

\begin{figure*}
\includegraphics[width=0.49\hsize]{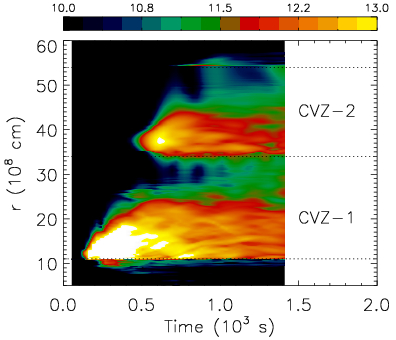} 
\includegraphics[width=0.49\hsize]{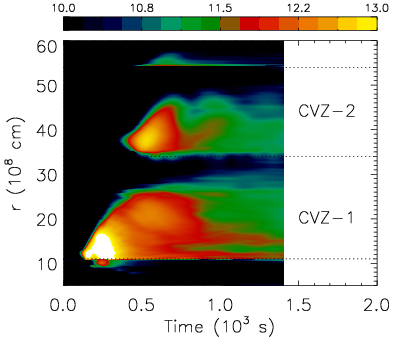} 
\caption{Temporal evolution of the radial distribution of the (color
  coded) logarithm of the angular averaged kinetic energy density (in
  \ergg) of models heflpopIII.2d.1 (left) and heflpopIII.3d (right),
  respectively. The horizontal dotted lines mark the boundaries of the
  double convection zone one part being sustained by the helium
  burning (CVZ-1) and the other one by the CNO cycle (CVZ-2).  }
\label{fig.kindnst}
\end{figure*}

Within the double convection zone originally determined by
  the Schwarzschild criterion the temperature gradient drops
  everywhere below the adiabatic one (Fig.\,\ref{fig.tmpgradpop3}).
  It does not imply that convection must cease. Even if the
  temperature gradient is not everywhere super-adiabatic nuclear burning
  may create hot blobs, which although cooling faster than the
  environment can still be hotter than the latter, and thus can rise
  upwards. 

This supports our conclusions based on the distribution of the
  Brunt-V\"ais\"al\"a frequency, which might be, however, a result of
  insufficient resolution, as the gradient increases with increasing
  resolution.  Consequently, we do not
  find the typical 2D convective pattern characterized by vortices in
  the double convection zone at later times t$~> 2000~s$.

\paragraph{Radiative barrier} 
%
Stellar evolutionary calculations of low-mass Pop III stars predict
that the helium flash-driven convection zone splits into two, when the
hydrogen-burning luminosity (driven by the entrainment of H) exceeds
the helium-burning luminosity. At this point, a radiative barrier is
created between both convective zones, as energy flows inwards from
the layers where hydrogen-burning takes place. The radiative barrier
is thought to prevent the flow of isotopes into the helium burning
layers and vice versa, hence preventing the reaction $^{13}$C
($\alpha$, n) $^{16}$O to become a source of neutrons. Also an
eventual mixing of isotopes from the helium-burning layer into the
stellar atmosphere should be inhibited.  However, this scenario is
difficult to prove due to numerical problems arising when modeling
this event in 1D \citep{Hollowell1990}.

Our hydrodynamic model heflpopIII.2d.2 with the double convective zone
shows that the radiative barrier allows for some interaction between
both zones via g-modes (Fig.\,\ref{fig.snap2d},
Fig.\,\ref{fig.kindnstdec}). In addition, there is some mixing of hydrogen 
into the radiative layer, which
was initially completely devoid of hydrogen (\ie X($^1$H) =
10$^{-30}$).  Hydrogen must have been mixed there either by convective
motion from the hydrogen-rich layers or dredged down by penetrating
convective plumes from the lower convection zone. The downward mixing
of hydrogen extends to a radius of $\sim 2.84\times 10^9\,$cm
(X($^1$H) $\sim 10^{-29}$) by the end of our simulation
heflpopIII.2d.2 (Fig.\,\ref{fig.hyddist}). It is likely that deeper
mixing of hydrogen into the helium burning layers is not possible,
since protons are captured via the reaction $^{12}$C (p, $\gamma$)
$^{13}$N on timescales shorter than that on which protons are mixed
inwards \citep{Hollowell1990}. At a temperature $T \sim 10^{8}\,$K ($r
\sim 2.21 \times 10^9\,$cm) the proton lifetime against capture by
$^{12}$C is as short as $\sim 10^2\,$s \citep{Caughlan1988}. This is
an order of magnitude smaller than the observed initial convective turnover
timescales (Tab.\,\ref{modpop3tab}).

\subsubsection{Simulation heflpopIII.2d.1 and heflpopIII.3d}
\label{subsect.2d3d}

We now discuss the qualitative behavior of 2D and 3D Pop III models
which were simulated using the same number of radial and angular
zones, but which have a lower radial and angular resolution than the
2D model heflpopIII.2d.2 discussed in the previous subsection.  The
quantitative properties of the convection zone of these models will
obviously be different as an increased grid resolution implies less
numerical viscosity and larger Reynolds numbers. We again stress here
that the characteristic Reynolds numbers in our 2D and 3D simulations
(Tab.\,\ref{modpop3tab}) are still many orders of magnitude smaller
than the values predicted by theory (see Sect.\,\ref{sect:regime}).

The comparison between the 2D and 3D simulations, heflpopIII.2d.1 and
heflpopIII.3d, provides important information on the impact of the
symmetry restriction imposed in the 2D models.  Contrary to the 2D
models, our 3D hydrodynamic simulations of turbulent flow performed
with the PPM scheme (Sect.\,\ref{sect:hcode}) are geometrically
unconstrained, \ie in the inertial regime turbulent eddies can decay
along the Kolmogorov cascade down to the finest resolved scales
\citep{Sitine2000}.

%
%
%

Due to the large computational cost we evolved the 3D model
heflpopIII.3d and the corresponding 2D model heflpopIII.2d.1 for
$0.39\,$hrs of stellar life, only. We find the following qualitative
differences between the 3D and 2D model (Fig.\,\ref{fig.tmpevol}): (i)
in 3D convection starts earlier in the outer part of the double
convection zone, (ii) convective velocities are smaller there, and
(iii) the convective structures have a plume-like shape in 3D 
(Fig.\,\ref{fig.snap3d}) and are
vortex-like in 2D (Fig.\,\ref{fig.snap2d}). On the other hand, the
models also exhibit the following common qualitative evolutionary
properties (Fig.\,\ref{fig.tmpevol}): (i) a decrease of the total
nuclear energy production rate, (ii) a decrease of the maximum
temperature, (iii) a decay of the velocity field in the convection
zones (Fig.\,\ref{fig.kindnst}), and (iv) the presence of internal 
gravity waves in the radiative barrier.

The differences observed between the 2D and 3D simulation do not come
at a surprise, as it is well known that 2D simulations lead to an
overestimate of the flow velocities \citep{Muthsam1995,Meakin2006}. On
the other hand, the common properties of the 2D and 3D model are also
shared by the high resolution simulation heflpopIII.2d.2, except for
the nuclear energy production rate, which does not decrease after
convection is fully established. This implies that both our 2D model
heflpopIII.2d.1 and 3D model heflpopIII.3d are not sufficiently well
resolved, although they show the most important characteristics of the
high resolution model heflpopIII.2d.2 described in
Sect.\,\ref{subsub:sim2d2}, \ie the presence of a decaying convective
flow in both convection zones which are later dominated by internal
gravity waves. This also holds for the intermediate radiative layer.

Contrary to the low resolution 2D model heflpopIII.2d.1, the
convective velocities  found for the 3D model heflpopIII.3d
\footnote{The convective velocities are measured just after convection
  appears for the first time during the simulation, as the convective
  flow decays very fast later.}
in the inner convection zone sustained by helium burning match those
predicted by stellar evolutionary calculations relatively well, although 
the modulus of the velocity is about a factor of two larger. In the outer 
part of the convection zone, sustained by the CNO cycle, the modulus of 
the velocity and the individual velocity components are smaller by more 
than a factor of two.
 
Interestingly, we find convection to be triggered spontaneously in
these simulations -- even without nuclear burning. This is highlighted
by the fact that the temporal evolution of the kinetic energy of the
3D model heflpopIII.3d with no nuclear nuclear energy production is
almost identical to that of the corresponding model with burning
switched on (Fig.\,\ref{fig.tmpevol}). Thus, we conclude that the
hydrodynamic convective flow observed in our models is mainly driven
by the adopted temperature gradient which is inherited from the 1D
stellar model, and is only partially sustained by nuclear burning
within the hydrodynamic simulation.

\section{Summary}
\label{sect:sum}

We have performed and analyzed a 3D hydrodynamic simulation of a core
helium flash near its peak in a Pop I star possessing a single
convection zone (single flash) sustained by helium burning. The
simulation covers 27\,hrs of stellar life, or roughly 100 convective
turnover timescales. In addition, we performed and analyzed 2D and 3D
simulations of the core helium flash near its peak in a Pop III star
which has a double convection zone (dual flash) sustained by helium
and CNO burning, respectively. These simulations cover only 1.8\,hrs
and 0.39\,hrs of stellar life, respectively, as convection dies out
shortly after it appears.

The convective velocities in our hydrodynamic simulation of the single
flash model and those predicted by stellar evolutionary calculations
agree approximately. Contrary to our previous findings, the
temperature gradient in the convection zone remains superadiabatic,
probably because of the increased spatial resolution of these
simulations as compared to our old models. As expected, the simulation
shows that the convection zone grows on a dynamic timescale due to
turbulent entrainment. This growth can lead to hydrogen injection into
the helium core as predicted by stellar evolutionary calculation of
extremely metal-poor or metal-free Pop III stars. Hydrogen injection
leads to a split of the single convection zone into two parts
separated by a supposedly impenetrable radiative zone. 
Our hydrodynamic simulations of the double convection zone show that
the two zones vanish as their convective motion decays very
fast. However, this result may be caused by an insufficient spatial
grid resolution or probably because the conditions 
represented by the stabilized initial model are a bit different from 
those of the original stellar model. While the convective velocities in our 2D
hydrodynamic models do not match those predicted by stellar
evolutionary calculations for the double convection zone at all, a
rough agreement is found in our 3D model for the velocities in the
inner convection zone sustained by helium burning.

%
\begin{acknowledgements}
The simulations were performed at the Leibniz-Rechenzentrum of the
Bavarian Academy of Sciences \& Humanities on the SGI Altix 4700
system.  The authors want to thank Frank Timmes for some of his
publicly available Fortran subroutines which we used in the Herakles
code. Miroslav Moc\'ak acknowledges financial support from the
Communauté francaise de Belgique - Actions de Recherche Concertées.
SWC acknowledges the support of the Consejo Superior de Investigaciones 
Científicas (CSIC, Spain) JAE-DOC postdoctoral grant and the MICINN grant 
AYA2007-66256. Part of this study utilised the Australian National Facility 
supercomputers, under Project Code g61.
\end{acknowledgements}

\bibliography{referenc}

\end{document}